\documentclass[preprint, 10pt]{elsarticle}
\usepackage[english]{babel}
\usepackage{amssymb}
\usepackage{mathrsfs}
\usepackage{multirow, hhline}   
\usepackage[table,xcdraw]{xcolor}
\usepackage{soul}
\usepackage{etoolbox}
\robustify{\cite}

\usepackage{amsmath}
\allowdisplaybreaks[4]
\usepackage{caption}
\usepackage{subcaption}
\usepackage{a4wide}
\usepackage{booktabs}   
\usepackage{array}      
\usepackage{siunitx}
\usepackage{hyperref}
\hypersetup{hidelinks,
        pdftitle={},    
        colorlinks=true,      
        linkcolor=blue,         
        citecolor=blue,     
        filecolor=blue,     
        urlcolor=blue,
	breaklinks=true}
    
\usepackage{enumitem}
\usepackage{algorithm}
\usepackage{algpseudocode}
\usepackage{cleveref}

\usepackage{geometry}
\usepackage{lineno}

\usepackage[skip=5pt]{subcaption}

\usepackage{doi}  
\usepackage{color}

\usepackage{pifont}

\journal{arXiv}

\begin{document}
	
\begin{frontmatter}

\title{Active learning with physics-informed neural networks for optimal sensor placement in deep tunneling through transversely isotropic elastic rocks}

\author[Cornell]{Alec Tristani}
\author[Cornell]{Chloé Arson}

\cortext[cor]{Corresponding author}
\fntext[fncor]{email: ayt34@cornell.edu}

\address[Cornell]{Cornell University, Department of Earth and Atmospheric Sciences, Ithaca, NY 14850, USA}

\begin{abstract}
This paper presents a deep learning strategy to simultaneously solve Partial Differential Equations (PDEs) and back-calculate their parameters in the context of deep tunnel excavation. A Physics-Informed Neural Network (PINN) model is trained with synthetic data that emulates in-situ displacement measurements in the host rock and at the cavity wall, obtained from extensometers and convergence monitoring. As acquiring field observations can be costly, a sequential training approach based on active learning is implemented to determine the most informative locations for new sensors. In particular, Monte Carlo dropout is used to quantify epistemic uncertainty and query measurements in regions where the model is least confident. This approach reduces the amount of required field data and optimizes sensor placement. The PINN is tested to reconstruct the displacement field around a deep tunnel of circular section excavated in transversely isotropic elastic rock and to determine the rock constitutive and stress-field parameters. Results demonstrate excellent performance on small, scattered, and noisy datasets, achieving high precision for the Young's moduli, shear modulus, horizontal-to-vertical far-field stress ratio, and the orientation of the bedding planes. The proposed framework is compatible with traditional observational methods and shall ultimately support decision-making for optimal subsurface monitoring and for near real-time adaptive tunnel design and control.
\end{abstract}

\begin{keyword}
Active Learning
\sep Physics-informed Neural Networks
\sep Inverse Analysis
\sep Deep Tunnels
\sep Transversely Isotropic Elastic Rocks
\sep Optimal Sensor Placement

\end{keyword}

\end{frontmatter}

\section{Introduction}\label{sec:Intro}

The ground response to tunnelling is usually simulated using computational models. Accurate determination of the governing physical parameters is therefore crucial for the robust design and long-term performance of underground infrastructure. This often necessitates field observations acquired during tunnel excavation as they provide information about the behavior at tunnel scale \cite{boidy_back_2002, sterpi_visco-plastic_2009, pellet_contact_nodate}. As formalized by Peck in 1969 \cite{peck1969advantages} in the context of observational methods, computational models and their parameters are then considered as working hypotheses, subject to confirmation or modification during excavation. In that approach, field measurements, such as displacements, are used to iteratively update model parameters, ensuring an accurate simulation of the observed ground behavior. Observational methods help reduce construction time and costs, while enhancing safety and long-term performance. Nevertheless, their successful application requires two conditions: (i) observations must be reliable and reveal significant physical phenomena; (ii) back-analyses performed to update model parameters from field measurements must be computationally efficient.

As each rock constitutive relationship yields a different model formulation, performing back-analyses requires specific computational implementations. For instance, Sakurai and Takeuchi \cite{sakurai_back_1983} developed a finite element (FE) formulation in which the rock behavior was assumed linear elastic to estimate the Young’s modulus of the ground, using displacement measurements obtained at the cavity wall (from convergence instruments) and within the rock mass (from extensometers). When closed-form solutions are available, least-squares optimization methods are suitable. For example, Lecampion et al. \cite{lecampion_parameter_2002} used synthetic data that emulates extensometer measurements to obtain the constitutive parameters of a Perzyna elasto-visco-plastic model. Schoen et al. \cite{schoen2022application} developed a three dimensional FE numerical model to perform inverse analysis from settlement measurements applied to mechanized tunnelling. Tristani et al. \cite{tristani_analytical_2024} derived an analytical solution for tunnels excavated in fractional viscoelastic plastic rocks and implemented a least-squares optimization scheme to characterize the rock mass. However, when the model is computationally costly, performing back-analysis of field data may be time-consuming as numerous forward passes are necessary. Efficient computational methods are therefore required.

Recent advances in differential programming (DP) have opened new avenues for the simulation of physical models \cite{innes2019differentiableprogrammingbridgemachine, rackauckas2021universaldifferentialequationsscientific, blondel2025elementsdifferentiableprogramming,jaxmat}, where several classes of methods have been developed for solving both forward and inverse problems. By leveraging automatic differentiation (AD) \cite{baydin2018automatic}, DP allows efficient and exact gradient computation (at machine precision) with respect to model inputs and parameters in an end-to-end manner, avoiding the need for manually derived adjoint formulations. Within this framework, both discrete and continuous formulations have been developed. While discrete approaches, whether based on FE formulations \cite{Xue_2023} or material point methods \cite{du2025jaxmpmlearningaugmenteddifferentiablemeshfree} solve the Partial Differential Equations (PDE) of the problem over a discretized mesh, continuous approaches approximate the solution with parametric functions over the entire domain. This continuous representation facilitates the evaluation of the solution at arbitrary locations, which can be advantageous when observations are sparse and irregularly distributed, as often encountered in tunnelling. In the context of Physics-Informed Machine Learning (PIML), Physics-Informed Neural Networks (PINNs) \cite{raissi_physics-informed_2019} approximate the solution in a continuous manner using neural networks, without relying on an explicit mesh. The governing equations are then enforced by evaluating PDE residuals at collocation points using AD.

More specifically, by seamlessly integrating prior knowledge from observational data and from PDEs, PINNs can make accurate predictions with small datasets, while other traditional deep learning techniques require large amounts of data \cite{karniadakis_physics-informed_2021}. In the PINN framework, the inverse problem is solved by simultaneously optimizing the network and physical parameters within a single unified loss function that combines PDE residuals, boundary condition residuals, and data misfit terms. PINNs have already been successfully applied across a wide range of scientific domains \cite{cuomo_scientific_2022} as an alternative to numerical methods based on discretization schemes. In fluid mechanics, they have shown particular effectiveness \cite{cai_physics-informed_2021}, for example for solving inverse problems of unsaturated groundwater flow \cite{depina_application_2022}. In solid mechanics, PINNs have been employed to solve both forward and inverse problems in elastic and plastic materials \cite{haghighat_physics-informed_2021}, as well as to predict dynamic stress fields \cite{bolandi_physics_2023}. In tunneling, PINNs have been used to predict ground surface settlements and estimate geotechnical parameters from settlement displacement data \cite{zhang_physics-informed_2023, cai_physics-informed_2025, wang_physics-data-driven_2025} or to estimate tunnel lining loads \cite{xu_transfer_2023, wang_estimation_2024}.

Despite such promising achievements, the performance of PINNs strongly depends on the amount and distribution of the training data, including both observational measurements and residual (collocation) points. When sufficient observations cannot be obtained, one remedy is to train Machine Learning (ML) models on smaller but more informative datasets that capture the essential complexities of the underlying mapping, which can be facilitated by data sampling strategies \cite{fuhg_review_2025}. When the number of measurements is fixed, adaptive sampling of residual points has proven to improve PINN accuracy. For instance, Lu et al. \cite{lu_deepxde_2021} suggested adding collocation points in regions exhibiting high PDE residuals. Nabian et al. \cite{nabian_efficient_2021} introduced an importance sampling algorithm to select the residual training points that contribute the most to the loss function. Wu et al. \cite{wu_comprehensive_2023} compared uniform and non-uniform sampling strategies and developed a refined selection technique to minimize residual errors. However, in all the aforementioned studies, the number of field measurements is assumed to be fixed, and only the distribution of collocation points is optimized. As the inverse problem is often ill-posed and sensitive to the location and quantity of observation data \cite{tarantola2005inverse}, one may therefore need to query the most informative field observations in order to reveal the significant phenomena and identify the governing physical parameters \cite{schoen2022application}.

This can be achieved through the computational framework provided by modern Optimal Experimental Design (OED) \cite{huan_optimal_2024}. Particularly, in the context of PIML, Active Learning (AL) relates to a class of algorithms designed to optimally choose the data to label and focuses on learning supervised predictive models \cite{dasgupta_two_2011}. The main idea behind AL is that a machine learning algorithm can achieve greater accuracy with fewer labeled training instances if it is allowed to choose the training data from which it learns \cite{settles_active_2009}. AL algorithms sequentially select one new data point to label at a time, within a data-pool. As a special case of OED, AL aims to improve downstream predictions by maximizing information gain rather than by calibrating model parameters \cite{rainforth_modern_2024}. An important point is to specify the criterion to select new data, for example, by labeling the data for which the uncertainties of the current model are the highest \cite{raissi_inferring_2017, sahli_costabal_physics-informed_2020, zhang_quantifying_2019, yang_b-pinns_2021}, by using query-by-committee techniques \cite{seung_query_1992, smith_less_2018}, or by minimizing the loss function residual \cite{arthurs_active_2021, mao_physics-informed_2023, gao_active_2023}. In most cases, AL outperforms traditional random sampling \cite{musekamp_active_2025}.

Thus, in this study, we leverage the computational paradigm of PINNs to both reconstruct the displacement field around a deep tunnel and determine the constitutive parameters of the rock mass from sparse and noisy measurements, while simultaneously optimizing the placement of convergence and extensometer sensors. To efficiently acquire new observations, an active learning strategy is implemented to sequentially query and label the measurements that are expected to most significantly improve model performance. To this end, the PINN is trained using dropout as a stochastic regularization technique \cite{srivastava2014dropout} which enables epistemic uncertainty estimation at inference time through Monte Carlo dropout \cite{gal_dropout_2016} and allows the model to query sensors in regions where predictions are least confident. The proposed approach aims to align with traditional observational methods and to support decision-making during the excavation phase by providing near real-time model parameter updates.

To illustrate the approach, we focus on a host rock that exhibits an elastic transversely isotropic behavior. In the field of tunneling, this constitutive model has been successfully used for grounds exhibiting a strong anisotropic behavior and for characterizing the instantaneous response of the rock mass. A synthetic dataset is generated using the high-fidelity finite element code MOOSE \cite{permann2020moose}. For such behaviors, it is worth emphasizing that analytical solutions to the problem of a cavity subjected to a biaxial far field stress have been derived in the past from the complex variable theory \cite{lekhnitskii_theory_1964}.  For example, the displacement field was expressed in the form of integer series for unlined \cite{hefny_analytical_1999} and lined \cite{bobet_lined_2011} circular tunnels, and for tunnels of any cross-sectional shape (using conformal mapping techniques) \cite{tran_manh_closed-form_2015}. Those analytical solutions assume that the constitutive parameters are known. Back-analysis approaches have also been proposed. For the cavity expansion problem, Kolymbas et al. \cite{kolymbas_cavity_2012} suggested an approximate solution to estimate the material constants, while Vu et al. \cite{vu_semi-analytical_2013} developed a semi-analytical solution to model displacement and stress fields around a tunnel section and to calculate constitutive parameters of the rock from convergence data. 

The remainder of this manuscript is organized as follows. Section \ref{sec:theoretical_background} presents some theoretical background on PINNs and introduces the excavation problem in transversely isotropic elastic grounds in the form of non-dimensional equations. Section \ref{sec:AL} details the model and formulates the active learning strategy developed in this study to best acquire the training data. Section \ref{sec:results} discusses the results obtained from a case study with several querying strategies, and demonstrates the effectiveness of the proposed approach using small, scattered, and noisy data. Finally, in Section \ref{sec:Conclusion}, conclusions are drawn, and perspectives for future research are presented.

\section{Theoretical background} \label{sec:theoretical_background}
\subsection{Physics-informed neural networks (PINNs)}
Let us consider the following generic PDE expressed as: 
\begin{equation}
\left\{
	\begin{aligned}
		\mathcal{D}\left[ \boldsymbol{u}\left( \boldsymbol{x} \right); \boldsymbol{\lambda}  \right] & = 0, \boldsymbol{x}\in \varOmega,
		\\
		\mathcal{B}\left[ \boldsymbol{u}\left( \boldsymbol{x} \right); \boldsymbol{\lambda} \right] & = 0, \boldsymbol{x}\in \partial \varOmega 
	\end{aligned}
    \right.
\label{eq:pde}
\end{equation}

Here, $\mathcal{D}\left[ \cdot \right]$ denotes a linear differential operator acting on $\boldsymbol{u}$, the solution of the differential equation; $\mathcal{B}\left[ \cdot \right]$ are boundary operators, $\boldsymbol{\lambda}$ denotes the PDE parameters, $\boldsymbol{x}$ is a position input vector in the spatial domain $\varOmega \subseteq \mathbb{R}^d$ having boundary $\partial \varOmega$, where $d \in \{1,2,3\}$. 

The PINN aims to approximate the true solution $\boldsymbol{u}(\boldsymbol{x})$ with a deep feedforward neural network $\mathcal{N}(\boldsymbol{x};\boldsymbol{\theta})$, where $\boldsymbol{\theta} = \{\boldsymbol{w}, \boldsymbol{b}\}$ denotes the trainable weights and biases. The network $\mathcal{N}(\boldsymbol{x};\boldsymbol{\theta})$ maps the spatial coordinates $\boldsymbol{x}$ to an approximation of the solution via a sequence of nested layer-wise transformations:
\begin{equation}\label{eq:nn_composition}
    \boldsymbol{z^l} = f(\boldsymbol{w}^{l}\boldsymbol{z}^{l-1} + \boldsymbol{b}^{l}) \quad l=1, \dots , L
\end{equation}
where $\boldsymbol{z}^0=\boldsymbol{x}$ and $\boldsymbol{z}^L=\mathcal{N}(\boldsymbol{x};\boldsymbol{\theta}) \approx \boldsymbol{u}(\boldsymbol{x})$ and where $f$ corresponds to the activation function which carries the non-linearity of the system. Usually, $f=\tanh$ for all layers except the last one, which is linear. 

Neural networks are trained to minimize a loss function $\mathcal{L}$. For PINNs, this loss function is composed of three terms, as follows:
\begin{equation}\label{eq:loss_function}
    \mathcal{L}(\boldsymbol \theta) = w_{pde}\mathcal{L}_{pde}(\boldsymbol\theta) + w_{bc}\mathcal{L}_{bc}(\boldsymbol\theta)
    + w_{data}\mathcal{L}_{data}(\boldsymbol\theta)
\end{equation}
with:
\begin{equation}\label{eq:loss_terms}
\left\{
\begin{aligned}
        \mathcal{L}_{pde}(\boldsymbol \theta) & = \dfrac{1}{N_{\Omega}} \sum_{i=1}^{N_\Omega} \left\lVert \mathcal{D}\left[ \mathcal{N}(\boldsymbol{x_i};\boldsymbol{\theta}); \boldsymbol{\lambda}\right] \right\rVert^2\\
        \mathcal{L}_{bc}(\boldsymbol \theta) & = \dfrac{1}{N_{\partial\Omega}} \sum_{i=1}^{N_{\partial\Omega}} \left\lVert \mathcal{B}\left[ \mathcal{N}(\boldsymbol{x_i};\boldsymbol{\theta}); \boldsymbol{\lambda}\right] \right\rVert^2\\
        \mathcal{L}_{data}(\boldsymbol \theta) & = \dfrac{1}{N_{data}} \sum_{i=1}^{N_{data}} \left\lVert \mathcal{N}(\boldsymbol{x_i};\boldsymbol{\theta}) - \boldsymbol{\hat{y}_i}\right\rVert^2\\
\end{aligned}
\right.
\end{equation}
where $\mathcal{L}_{pde}(\boldsymbol{\theta})$ represents the residual of the PINN solution, i.e., the extent to which the predicted solution fails to satisfy the governing PDEs that encode the physical constraints; $\mathcal{L}_{bc}(\boldsymbol{\theta})$ represents the deviation from the boundary conditions (BCs); and $\mathcal{L}_{data}(\boldsymbol{\theta})$ represents the discrepancy between the PINN-predicted solution and the available data at observation points. 

In equation \ref{eq:loss_terms}, $\{\boldsymbol{x_i}\}_{i=1}^{N_\Omega}$ are residual points (also called collocation points) located in the domain $\Omega$; $\{\boldsymbol{x_i}\}_{i=1}^{N_{\partial\Omega}}$ are boundary points; and $\{\boldsymbol{x_i}, \boldsymbol{\hat{y}_i}\}_{i=1}^{N_{data}}$ are measurements. $N_\Omega, N_{\partial\Omega}, N_{data}$ denote the total number of collocation points, boundary points, and data points, respectively. $w_{pde}$, $w_{bc}$ and $w_{data}$ are the weights associated with the PDE loss, BC loss, and data loss, respectively.

The network parameters $\boldsymbol\theta$ are tuned by minimizing the total training loss $\mathcal{L}(\theta)$. This is done during back-propagation at machine precision, i.e., at the highest numerical precision allowed by the floating-point representation on the hardware (typically float32 or float64), by leveraging AD \cite{baydin2018automatic} and using the Leibniz chain rule. In this study, AD is implemented in PyTorch \cite{paszke_pytorch_2019}. Different optimizers can be chosen: stochastic optimization schemes such as Adam \cite{kingma_adam_2014} or Nadam \cite{dozat_incorporating_2016}, or deterministic optimizers such as L-BFGS \cite{liu_limited_1989}, which have also been widely used to train PINNs. A hybrid approach combining both optimizers can be used effectively to improve convergence \cite{wu_comprehensive_2023}. In this study, the stochastic Nadam optimizer is used for training, while dropout \cite{gal_dropout_2016} is further leveraged to train the model (see Section \ref{sec:sequential_training}). Note that the weights $w_{data}, w_{pde}, w_{bc}$ are usually set constant prior to minimizing the total loss, which is the case in this work. They can alternatively be adapted during training, as proposed in the Self-Adaptive PINN method \cite{mcclenny_self-adaptive_2023} or based on learning rate annealing methods \cite{wang_understanding_2020}. In this work, fixed weights are assigned to each loss term as detailed in Section \ref{sec:model}.

PINNs can be used to solve both forward and inverse problems. In the forward problem, the governing PDEs, material parameters, and boundary conditions are known and fixed. The objective is to compute the solution field $\boldsymbol{u}$. The PINN is trained to minimize the PDE residuals and to satisfy boundary conditions, and no observational data is required. Note that in the case of forward problems, energy-based formulations have also been proposed for elasticity and hyperelasticity problems \cite{samaniego2020energy, nguyen2020deep, fuhg2022mixed} as well as for viscoelasticity \cite{abueidda2022deep}. These approaches present advantages such as requiring only first-order derivatives and naturally satisfying homogeneous Neumann boundary conditions but are less suited for solving the inverse problem \cite{thombre2025energy}.

In the inverse problem, some coefficients of the PDEs or related to the boundary/initial conditions are unknown. The objective is to identify those unknown parameters from scattered and noisy measurements. The PINN is therefore trained to satisfy both the PDE/BC constraints and to minimize the data loss simultaneously. In practice, the unknown parameters are treated as additional trainable variables within the parameter set $\boldsymbol{\theta}$. Inverse problems are often ill-posed in the sense of Hadamard: a solution may not exist, may not be unique, or may be highly sensitive to the observational data. In the following, we focus on solving the inverse problem.

\subsection{Physical governing equations}
Rocks naturally exhibit heterogeneities. When the scale of those heterogeneities is similar to that of the tunnel diameter, the host rock cannot be considered as a continuum, and field discontinuities must be embedded in the formulation of the problem. Here, we study instead anisotropic rock formations that present heterogeneities that are at least two orders of magnitude smaller than the dimensions of the tunnel cross-section. We focus on sedimentary or metamorphic rocks, which develop a layered structure (through deposition of grains and particles in successive layers in the case of sedimentary rocks, or foliation during metamorphism in metamorphic rocks). This anisotropy can be accounted for through a transverse isotropic elastic behavior model to describe the response of the tunnel \cite{hefny_analytical_1999, tonon_effect_2002, kolymbas_cavity_2012, tran_manh_closed-form_2015}. In this case, the elasticity tensor depends on $5$ independent parameters \cite{noauthor_elasticity_2006}.

\begin{figure}[h]
    \centering
    \begin{subfigure}{0.43\linewidth}
        \centering
        \includegraphics[scale=1]{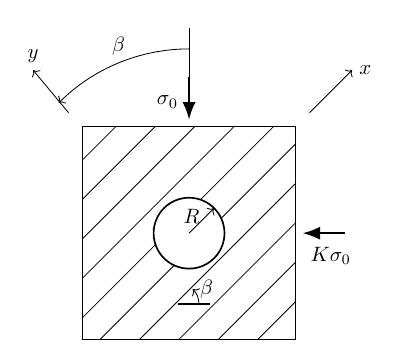}
        \centering
        \caption{In-situ loading conditions}
        \label{fig:in_situ_condition}
    \end{subfigure}
    \centering
    \begin{subfigure}{0.43\linewidth}
        \centering
        \includegraphics[scale=1]{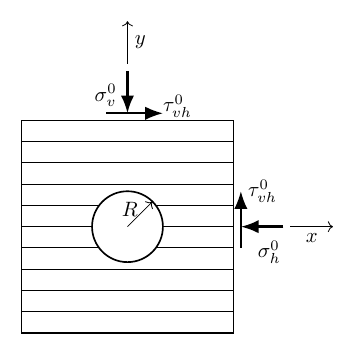}
        \centering
        \caption{Loading after rotation}
        \label{fig:loading_after_rotation}
    \end{subfigure}
    \caption{Tunnel with circular cross-section of radius $R$ excavated in transversely isotropic elastic rocks. }
    \label{fig:tunnel_pb}
\end{figure}

In the following, we consider a deep circular tunnel excavated in a transversely isotropic elastic ground. We define the $x$-axis and $z$-axis as the axes that define the plane of isotropy of the host rock, as shown in Figure \ref{fig:in_situ_condition}. In the (x,y) plane, the angle formed between the bedding direction (x-axis) and the horizontal is noted $\beta$. The tunnel under study is excavated along the $z$-axis, which corresponds to the most unfavorable case, since the maximum deformations are expected to occur in the $y$-direction, which is the direction normal to the bedding plane. Two-dimensional plane-strain conditions are assumed. The in-situ stress state in a plane orthogonal to the $z$-axis is assumed to be homogeneous and anisotropic. Taking the horizontal and vertical axes as first and second directions of space, respectively, the in-situ stress state is defined as:
\begin{equation}\label{eq:init_stress}
    \boldsymbol{\sigma}^i = 
    \begin{pmatrix}
        K\sigma_0 & 0 \\
        0 & \sigma_0
    \end{pmatrix}
\end{equation}
where $K$ is the ratio between the horizontal and vertical stress (lateral earth pressure coefficient) and $\sigma_0$ is the far field vertical stress. A rotation by $2\beta$ in the stress space ($\beta$ in the geometric space) yields the in-situ stress state components in the (x,y) coordinate system (see Figure \ref{fig:loading_after_rotation}), as follows:
\begin{equation}\label{eq:init_stress_rot}
\left\{
\begin{aligned}
\sigma_h^0 & = \dfrac{\sigma_0}{2}\left[1+K - (1-K)\cos(2\beta) \right]\\
\sigma_v^0 & = \dfrac{\sigma_0}{2}\left[1+K + (1-K)\cos(2\beta) \right] \\
\tau_{vh}^0 & = \dfrac{\sigma_0}{2}(K-1)\sin(2\beta)\\
\end{aligned}
\right.
\end{equation}

The total deformation $\boldsymbol\varepsilon^t$ is decomposed into two parts: the so-called eigenstrain $\boldsymbol\varepsilon^{i}$, which corresponds to the initial deformation of the ground due to the in-situ state of stress; and $\boldsymbol\varepsilon$, the deformation of the ground due to the excavation of the tunnel, which redistributes stresses around the cavity. The excavation deformation $\boldsymbol\varepsilon$ is thus calculated as:

\begin{equation}\label{eq:exc_init_strain}
    \boldsymbol\varepsilon = \boldsymbol{\varepsilon^{t}} - \boldsymbol{\varepsilon^{i}}
\end{equation}
Noting $\boldsymbol{S}$ the compliance tensor of the transverse anisotropic rock mass, the stress and strain fields satisfy:
\begin{equation}\label{eq:eigenstrains}
    \boldsymbol{\varepsilon^i} = \boldsymbol{S}:\boldsymbol{\sigma^i}
\end{equation}
and 
\begin{equation}\label{eq:el}
    \boldsymbol{\varepsilon^t} = \boldsymbol{S}:\boldsymbol{\sigma}
\end{equation}
where $\boldsymbol{\sigma}$ is the Cauchy stress tensor which corresponds to the stress field around the cavity.

In plane-strain, using Voigt notation for equation \ref{eq:exc_init_strain} yields:
\begin{equation}\label{eq:constitutive_law}
\begin{pmatrix}
    \varepsilon_{xx} \\
    \varepsilon_{yy} \\
    \gamma_{xy}
\end{pmatrix}
=
\begin{pmatrix}
\displaystyle \frac{1-\nu_h^2}{E_h} & \displaystyle -\frac{\nu_{vh}(1+\nu_h)}{E_v} & 0  \\[8pt]
\displaystyle -\frac{\nu_{vh}(1+\nu_h)}{E_v} & \displaystyle \frac{1-\nu_{vh}\nu_{hv}}{E_v} & 0 \\[8pt]
0 & 0 & \displaystyle \frac{1}{G_{vh}} \\[8pt]
\end{pmatrix}
\begin{pmatrix}
    \sigma_{xx} - \sigma_h^0 \\ 
    \sigma_{yy} - \sigma_v^0 \\
    \tau_{xy} - \tau_{vh}^0
\end{pmatrix}
\end{equation}
where $E_h$ and $E_v$ denote the Young's moduli in the plane of isotropy and in the direction normal to that plane, respectively. $\nu_h$ is the Poisson's ratio for deformation within the plane of isotropy. $\nu_{hv}$ and $\nu_{vh}$ represent, respectively, the Poisson's ratio for the effect of in-plane stress on the strain normal to the plane, and the Poisson's ratio for the effect of normal stress on the strain within the plane of isotropy. These parameters are related by the reciprocity condition:
\begin{equation}
    \nu_{hv} E_v = \nu_{vh} E_h.
\end{equation}
Lastly, $G_{vh}$ is the shear modulus in any plane containing the normal direction to the plane of isotropy.

For small strains, the strain tensor is given by:
\begin{equation}\label{eq:small_strains}
    \boldsymbol{\varepsilon} = \frac{1}{2} \left[ \boldsymbol{\nabla }\boldsymbol{u} + (\boldsymbol{\nabla} \boldsymbol{u})^{T} \right],
\end{equation}
where $\boldsymbol{u}(\boldsymbol{x}) = (u_x,u_y)$ is the displacement vector at position $\boldsymbol{x}$. In the absence of body forces, the equilibrium equation is:
\begin{equation}
    \label{eq:equilibrium}
    \nabla \cdot \boldsymbol{\sigma} = \boldsymbol{0}
\end{equation}

The boundary condition at the tunnel wall is:
\begin{equation}\label{eq:bc}
    \boldsymbol{\sigma} \cdot \boldsymbol{n} = \boldsymbol{0}
\end{equation}
where $\boldsymbol{n}$ is the vector normal to the cavity wall. Note that this condition corresponds to the case of a tunnel that is fully excavated. Far away from the cavity, the perturbation induced by the excavation vanishes. Thus, the displacement field satisfies the far-field boundary condition:
\begin{equation}\label{eq:bc_far}
\lim_{\lVert \boldsymbol{x} \rVert \to \infty} \boldsymbol{u}(\mathbf{x}) = \boldsymbol{0}.
\end{equation}

\subsection{Non-dimensionalization}

To avoid vanishing or exploding gradients during back-propagation, it is good practice to scale input and output variables. In the following, the domain is bounded by the tunnel and the locations where measurements are collected. We note $R$ the radius of the tunnel section, and $L$ the length of any extensometer that could be placed around the cavity to collect measurements. As suggested in \cite{xu_transfer_2023}, we introduce the following dimensionless variables:
\begin{equation}\label{eq:dimensionless_variables}
    \Tilde{x} = \dfrac{x}{l_a}, \quad 
    \Tilde{y} = \dfrac{y}{l_a}, \quad 
    \Tilde{u}_x = \dfrac{u_x}{u_{a}} \quad
    \Tilde{u}_y = \dfrac{u_y}{u_{a}}
\end{equation}
where:
\begin{equation}\label{eq:scaling_factors}
    l_a = R + L, \quad 
    u_a = 10^{-1} m, \quad \text{and}\quad
    \sigma_a = \sigma_0 \dfrac{l_a}{u_{a}}.
\end{equation}
This normalization serves two purposes: it provides a dimensionless formulation of the governing equations and ensures that the input and output variables remain within numerically convenient ranges during training. Note that the scaling coefficient $u_a$ was set to $ 10^{-1}$ m because the maximum displacement expected from the numerical analysis does not exceed $10^{-1}$ m. 

Inserting the dimensionless variables defined in equation \(\ref{eq:dimensionless_variables}\) into the equilibrium equation \(\ref{eq:equilibrium}\) yields the following expanded relationships:
\begin{equation}\label{eq:equilibrium_expanded_dimensionless}
\left\{
    \begin{aligned}
    D\left[ \nu_{hv} (1 + \nu_h) \dfrac{\partial^2 \Tilde{u}_y}{\partial \Tilde{x} \partial \Tilde{y}
} + \left( \dfrac{\Tilde{E}_h}{\Tilde{E}_v} - \nu_{hv}^2 \right) \dfrac{\partial^2 \Tilde{u}_x}{\partial \Tilde{x}^2} \right] + \Tilde{G}_{vh} \left( \dfrac{\partial^2 \Tilde{u}_x}{\partial \Tilde{y}^2} + \frac{\partial^2 \Tilde{u}_y}{\partial \Tilde{x} \partial \Tilde{y}
} \right) = 0 \\
D \left[ \nu_{hv} (1 + \nu_h) \frac{\partial^2 \Tilde{u}_x}{\partial \Tilde{x} \partial \Tilde{y}
} + (1 - \nu_h^2) \frac{\partial^2 \Tilde{u}_y}{\partial \Tilde{y}
^2} \right] + \Tilde{G}_{vh} \left( \frac{\partial^2 \Tilde{u}_x}{\partial \Tilde{x} \partial \Tilde{y}} + \frac{\partial^2 \Tilde{u}_y}{\partial \Tilde{x}^2} \right) = 0  
    \end{aligned}
    \right.
\end{equation}
and inserting the dimensionless variables into the boundary equation \ref{eq:bc}, one obtains:
\begin{equation}\label{eq:bc_expanded_dimensionless}
\left\{
    \begin{aligned}
    \Tilde{x} D\left[ \nu_{hv} (1 + \nu_h) \dfrac{\partial \Tilde{u}_y}{\partial \Tilde{y}} + \left( \dfrac{\Tilde{E}_h}{\Tilde{E}_v} - \nu_{hv}^2 \right) \dfrac{\partial \Tilde{u}_x}{\partial \Tilde{x}} \right] + \Tilde{y} \Tilde{G}_{vh} \left( \dfrac{\partial \Tilde{u}_x}{\partial \Tilde{y}} + \frac{\partial \Tilde{u}_y}{\partial \Tilde{x}} \right) + \Tilde{x} \Tilde{\sigma}_h^0 + \Tilde{y} \Tilde{\tau}_{vh}^0 = 0 \\
  \Tilde{y} D \left[ \nu_{hv} (1 + \nu_h) \frac{\partial \Tilde{u}_x}{\partial \Tilde{x}} + (1 - \nu_h^2) \frac{\partial \Tilde{u}_y}{\partial \Tilde{y}} \right] + \Tilde{x} \Tilde{G}_{vh} \left( \frac{\partial \Tilde{u}_x}{ \partial \Tilde{y}} + \frac{\partial \Tilde{u}_y}{\partial \Tilde{x}} \right) + \Tilde{y} \Tilde{\sigma}_v^0 + \Tilde{x} \Tilde{\tau}_{vh}^0= 0        
    \end{aligned}
    \right.
\end{equation}
where:
\begin{equation}\label{eq:D}
    D = - \dfrac{\Tilde{E}_h}{2\nu_{hv}^2(1+\nu_h) - \dfrac{\Tilde{E}_h}{\Tilde{E}_v}(1-\nu_h^2)}
\end{equation}
with:
\begin{equation}\label{eq:dimensionless_parameters}
    \Tilde{E}_h = \dfrac{E_h}{\sigma_a}, \quad 
    \Tilde{E}_v = \dfrac{E_v}{\sigma_a}, \quad 
    \Tilde{G}_{vh} = \dfrac{G_{vh}}{\sigma_a}, \quad
    \Tilde{\sigma}_{h}^0 = \dfrac{\sigma_h^0}{\sigma_0} \quad
    \Tilde{\sigma}_{v}^0 = \dfrac{\sigma_v^0}{\sigma_0} \quad
    \Tilde{\tau}_{vh}^0 = \dfrac{\tau_{vh}^0}{\sigma_0}
\end{equation}

\section{Proposed approach: active learning to optimally place new sensors} \label{sec:AL}

\subsection{Problem setup}

During tunneling, assessing the response of the rock mass to excavation is essential for ensuring safety. As pointed out by Guayac\'an-Carrillo et al. \cite{guayacan2026lessons}, close monitoring of ground deformation around the tunnel provides useful information on ground response during and after excavation, as it helps reveal the significant observed phenomena. The collected measurements can subsequently serve as training data for machine learning models, enabling fast computations \cite{tristani_data_driven_2025, tristani_physics-informed_2025}. Monitoring the ground behavior can be carried out by installing extensometers and convergence instruments around the tunnel wall. However, this process may be both costly and time-consuming, and the placement of sensors is often guided by empirical approaches. Fundamental questions naturally arise: How many measurements are necessary? Where should the sensors be placed? And, given the available measurements, can we reconstruct the displacement field and identify the constitutive parameters of the rock mass? Addressing these questions would not only improve our understanding of in-situ phenomena, but also contribute to more efficient and reliable tunnel design \cite{schoen2022application}. To this end, we leverage PINNs to propose an automatic, data-efficient framework based on their predictive capabilities. Specifically, we aim to reconstruct the displacement field around the tunnel while simultaneously identifying the rock mass parameters, namely, the horizontal-to-vertical stress ratio and the orientation of the bedding planes, using the smallest possible amount of measurement data. 

Extensometers consist of radial cables installed around the tunnel cavity. They measure displacements at anchored points that are typically located every $3$–$4$ meters along a radial cable starting from the tunnel wall. Because the measurement reflects the relative displacement between the anchored points and the extensometer head, it may represent either a relative or absolute displacement depending on whether the head itself moves. In this study, we assume that the extensometer head remains fixed, so that the recorded displacements correspond to the true ground displacements. In addition to extensometer data, tunnel convergence is usually monitored. 

Convergence measurements correspond to the relative displacement between two opposite points of the tunnel wall at a given section. They can also be obtained as the three-dimensional displacements of individual wall points measured by all the convergence stations. Convergence data provide valuable information about tunnel closure, allowing for a rapid assessment of ground behavior and for design updates if needed \cite{sulem_closure_nodate}. Because convergence is the easiest measurement to obtain, it is commonly monitored during and after excavation. Convergence and extensometer measurements at the wall have been shown to display similar trends \cite{lara_time-dependent_2025}. Therefore, in this work, it is assumed that both convergence and extensometer measurements are equivalent indicators of tunnel closure.

\subsection{Synthetic dataset}

In the following, the measurements are assumed to be drawn from a pool of observations 
$\mathcal{P} = \{(x_i, y_i)\}_{i=1}^{N_{\text{data}}}$. These observations correspond to displacement measurements collected using field sensors. Each sensor, either an extensometer or a convergence point, provides an observation subset $\mathcal{S}_i \subset \mathcal{P}$. When referred to as an \textit{extensometer}, a sensor is represented by seven measurement points aligned along a radial direction within the domain and starting at the tunnel wall, with anchors placed every $4$ m. Consecutive extensometers are spaced by an angle of $10^\circ$. When referred to as a \textit{convergence point}, a sensor is a single measurement point located at the boundary of the domain, i.e., at the tunnel wall. For each sensor in the pool, in addition to computing the data loss term, the PINN also evaluates the PDE and BC residuals.

\begin{figure}[h]
    \centering
    \includegraphics[scale=1]{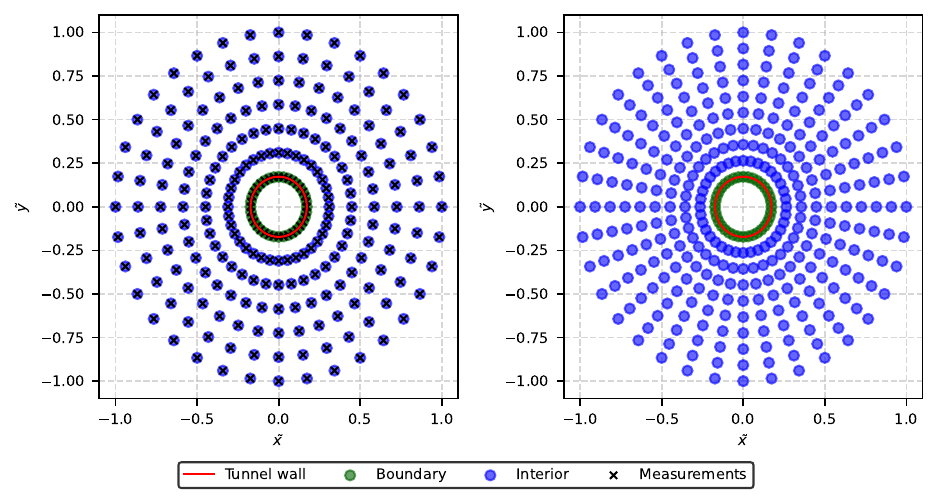}
    \caption{Training set. Left: pool of measurements $\mathcal{P}$ from which the model can query data, including both extensometers and convergence points. Right: grid of collocation and boundary points $\mathcal{G}$ used to solve the PDE and BC in the domain.}
    \label{fig:dataset}
\end{figure}

Additionally, a second grid $\mathcal{G} = \{(x_i, y_i)\}_{i=1}^{N_{\Omega}} \cup \{(x_i, y_i)\}_{i=1}^{N_{\partial \Omega}}$ of size $N_r \times N_\theta = 10 \times 36$ is defined, consisting of uniformly distributed collocation and boundary points. $\mathcal{G}$ is independent of the pool of observations $\mathcal{P}$ and is used throughout the training to enforce the PDE and BC in the domain. This ensures training stability and helps the model to accurately map the true solution around the tunnel. Therefore, the training set $\mathcal{T}$ consists of: (i) the $n$ sensors that the model has already queried $\{\mathcal{S}_i\}_{i=1}^{n}$, and (ii) the grid $\mathcal{G}$ containing the collocation and boundary points such that $\mathcal{T}= \{\mathcal{S}_i\}_{i=1}^{n} \cup \mathcal{G}$. The dataset is illustrated in Figure~\ref{fig:dataset}. 

The domain is scaled as described in Section~\ref{sec:theoretical_background}, so that all input coordinates and output displacements lie within the range $[-1, 1]$. Note that since we are solving an inverse problem, the displacement measurements collected near the cavity act as Dirichlet-type constraints on the solution. These local observations help constrain the model to identify the constitutive parameters within the region of interest. Points located far away from the cavity, which would correspond to prescribed Dirichlet conditions at the far-field, are not included in the dataset, so that the learning process focuses on the area where observations are available.

Lastly, model performance is evaluated on the full domain and along the boundary using a polar grid with $N_r = 50$ radial points and $N_\theta = 200$ angular points, both uniformly distributed, totaling $10{,}000$ evaluation points.

The reference dataset is generated using the high-fidelity finite element solver MOOSE. The computational domain is discretized using a mesh of 10,274 triangular linear elements, with local grid refinement near the tunnel wall. Collocation points are extracted using the LineValueSampler utility in MOOSE, which samples field values along predefined lines and interpolates when the target points do not coincide exactly with mesh nodes.

\subsection{PINN Model}\label{sec:model}

\begin{figure}[h]
    \centering
    \includegraphics[scale=1]{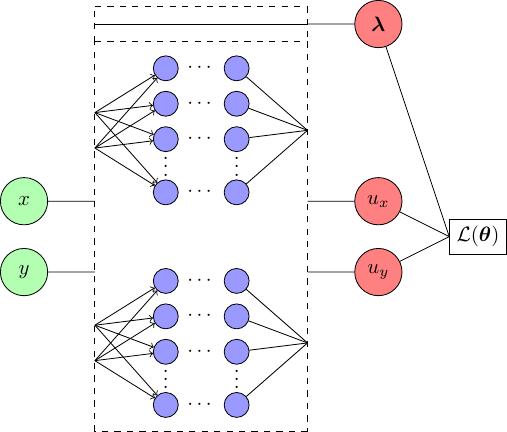}
    \caption{PINN architecture used in this study. Independent neural networks, $\mathcal{N}_{u_x}(\boldsymbol{x}; \boldsymbol{\theta})$ and $\mathcal{N}_{u_y}(\boldsymbol{x}; \boldsymbol{\theta})$, are defined to predict the displacement components $u_x$ and $u_y$, respectively. Each network takes the spatial coordinates $(x, y)$ as input features.}
    \label{fig:pinn_architecture}
\end{figure}

To improve training stability and accelerate convergence \cite{haghighat_physics-informed_2021}, independent neural networks are defined to predict each output variable, namely $u_x$ and $u_y$. In all experiments, the hyperbolic tangent (tanh) is used as the activation function. Each neural network consists of $10$ hidden layers, with $40$ neurons per layer. The displacement field is therefore approximated as:
\begin{equation}\label{eq:independent_nn}
\begin{aligned}
    u_x & \approx \mathcal{N}_{u_x}(\boldsymbol{x};\boldsymbol{\theta}) \\
    u_y & \approx \mathcal{N}_{u_y}(\boldsymbol{x};\boldsymbol{\theta})
\end{aligned}
\end{equation}
where $\mathcal{N}_{u_x}(\boldsymbol{x}; \boldsymbol{\theta})$ and $\mathcal{N}_{u_y}(\boldsymbol{x}; \boldsymbol{\theta})$ correspond to neural networks. This architecture is illustrated in Figure~\ref{fig:pinn_architecture}.

During training, the parameters of the PINN are optimized by the stochastic Nadam optimizer \cite{dozat_incorporating_2016}, which has been widely employed in the machine learning community. At each new iteration step, the model parameters, i.e., weights, biases, and PDE parameters, are initialized as those obtained at the last previous step. However, the optimizer is reinitialized, which means that its internal states are not retained across each step. This process avoids suboptimal solutions that correspond to local minima of the loss functions.

Equal weights can be assigned to each term of the loss function \cite{haghighat_physics-informed_2021}. However, we observed in this work that using different weights improved the performance of the PINN. In particular, attributing superior weights to the data and BCs losses compared to the PDE loss yielded better convergence as it helped physically constrain the model, an observation that has also been reported in previous studies \cite{wang_understanding_2020, wang_when_2022}. In the remainder of this work, the following weighting scheme is adopted:
\begin{equation}\label{eq:weights_scheme}
        w_{pde} = 1 \quad
        w_{bc} = 10 \quad
        w_{data} = 100 \quad
\end{equation}
A sensitivity analysis of the weighting scheme is presented in \ref{app:sensitivity}.

\subsection{Sequential learning approach}\label{sec:sequential_training}

In most machine learning approaches, the model passively relies on a fixed training dataset. In contrast, active learning enables the model to strategically select the most informative data points or measurements from which to learn. The fundamental objective is to achieve maximal predictive performance while minimizing the amount of data required, making active learning closely aligned with the principles of optimal experimental design in statistics \cite{settles_active_2009}. 

In this work, dropout is used to train the PINN. The advantages of using dropout are twofold: i) by randomly deactivating neurons during training, it enables the quantification of epistemic uncertainty through multiple inference passes, a process known as Monte Carlo (MC) dropout, which approximates Bayesian inference \cite{gal_dropout_2016}; and ii) it reduces overfitting to the training data, thereby improving model reliability \cite{srivastava2014dropout}.

A fixed dropout ratio equal to $5\%$ is used for the first two hidden layers. Monte Carlo dropout is then performed to assess epistemic uncertainties. The model then selects the sensors yielding the highest predictive uncertainty in a manner consistent with the active learning strategy proposed in \cite{raissi_inferring_2017, yang_b-pinns_2021}. Algorithm \ref{alg:query_strat} details how a new sensor is selected.

An Active Learning (AL) strategy is then designed to efficiently query the most informative data points within the domain and thereby improve the robustness and accuracy of the machine learning models, as described in Algorithm \ref{alg:active_learning}. Data points are queried sequentially, and the PINNs are trained iteratively in a stage-wise manner. New measurements are labeled from the pool of observations based on the epistemic uncertainties of the PINN, evaluated at the end of each training stage. Note that the approximate epistemic uncertainty estimated via MC Dropout serves here as a practical criterion for sensor selection, rather than as a comprehensive uncertainty quantification framework. The relevance of this criterion is then assessed through its effect on the accuracy of the reconstructed displacement fields, as presented in Section \ref{sec:results}.

\begin{algorithm}[H]
\caption{Query Strategy using Monte Carlo Dropout}
\begin{algorithmic}[1]
\State \textbf{Input:} trained model, candidate sensor set $\{\mathcal{S}_i\}_{i=1}^{N}$, current training set $\mathcal{T}$
\For{each candidate sensor $\mathcal{S}_i$}
    \State Perform $n_{MC}$ stochastic forward passes with dropout enabled $\mathcal{N}_{u_x}^{(k)}(\mathcal{S}_i), \mathcal{N}_{u_y}^{(k)}(\mathcal{S}_i), \quad k = 1, \dots, n_{MC}$ 
    \State Compute:
    \begin{itemize}
        \item mean prediction $\bar{u}_x$, $\bar{u}_y$  of $u_x, u_y$ over the $n_{MC}$ passes
        \item variance $\sigma_{u_x}^2$, $\sigma_{u_y}^2$ of $u_x, u_y$ over the $n_{MC}$ passes
    \end{itemize}
    \State Aggregate uncertainty of the sensor:
        $
            \bar{\sigma}^2(\mathcal{S}_i) = \text{mean of } (\sigma_{u_x}^2 + \sigma_{u_y}^2)
            \text{ over all points in } \mathcal{S}_i
        $
\EndFor
\State Select highest-uncertainty sensor:
    $
        \mathcal{S}^\star = \arg\max_{\mathcal{S}_i} \bar{\sigma}^2(\mathcal{S}_i)
    $
\State Update training set: $
    \mathcal{T} \gets \mathcal{T} \cup \mathcal{S}^\star
$
\State \textbf{return} updated training set $\mathcal{T}$
\end{algorithmic}
\label{alg:query_strat}
\end{algorithm}

\begin{algorithm}
\caption{Active Learning}
\begin{algorithmic}[1]
\State \textbf{Initialization:}
\State Select two initial extensometers $\mathcal{S}_1$, $\mathcal{S}_2$ from the pool of measurements $\mathcal{P}$ such that $\mathcal{T} = \mathcal{S}_1 \cup \mathcal{S}_2$.
\State Initialize optimizer
\State Train the PINN for $N$ epochs by computing $\mathcal{L}_{\text{pde}}$, $\mathcal{L}_{\text{bc}}$, and $\mathcal{L}_{\text{data}}$.
\State Add the grid $\mathcal{G}$: $\mathcal{T} \gets \mathcal{T} \cup \mathcal{G}$
\State Initialize optimizer
\State Train the PINN for $N$ epochs by computing $\mathcal{L}_{\text{pde}}$, $\mathcal{L}_{\text{bc}}$, and $\mathcal{L}_{\text{data}}$.
\Repeat
    \State Initialize optimizer
    \State Query next sensor as described in Algorithm \ref{alg:query_strat}
    \State Train the PINN for $N$ epochs by computing $\mathcal{L}_{\text{pde}}$, $\mathcal{L}_{\text{bc}}$, and $\mathcal{L}_{\text{data}}$.
    \State Update the training set: $\mathcal{T} \gets \mathcal{T} \cup \mathcal{S}$.
\Until{the maximum allowed number of sensors is reached.}
\end{algorithmic}
\label{alg:active_learning}
\end{algorithm}

\begin{figure}[H]
    \centering
    \begin{subfigure}{0.4\linewidth}
        \centering
        \includegraphics[scale=0.96]{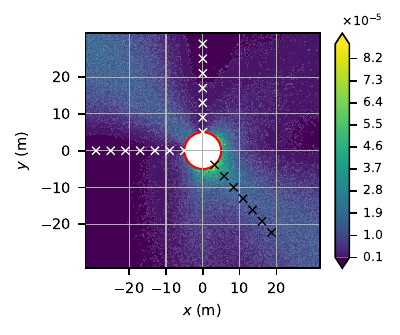}
        \centering
        \caption{Querying extensometers, Step 2}
    \end{subfigure}
    \centering
    \begin{subfigure}{0.4\linewidth}
        \centering
        \includegraphics[scale=0.96]{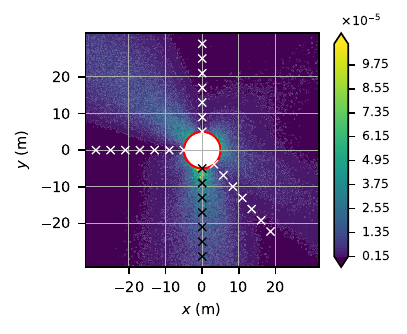}
        \centering
        \caption{Querying extensometers, Step 3}
    \end{subfigure}
    \begin{subfigure}{0.4\linewidth}
        \centering
        \includegraphics[scale=0.96]{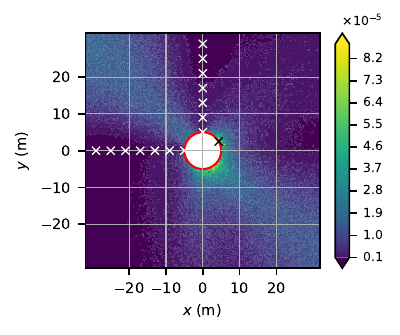}
        \centering
        \caption{Querying convergences, Step 2}
    \end{subfigure}
    \centering
    \begin{subfigure}{0.4\linewidth}
        \centering
        \includegraphics[scale=0.96]{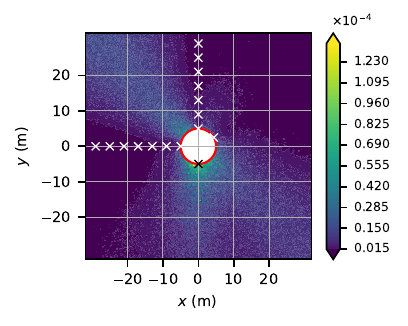}
        \centering
        \caption{Querying convergences, Step 3}
    \end{subfigure}
    \caption{Sequential active learning process where sensors are actively selected by the model and added to the training set. The color scale represents the total epistemic uncertainty $\sigma^2 = \sigma_{u_x}^2 + \sigma_{u_y}^2$. Current sensors are shown in white, and the next sensor to be queried is highlighted in black. (a)-(b) Active learning when only extensometer observation points are added sequentially. (c)-(d) Active learning when only convergence observation points are added sequentially.} 
    \label{fig:active_learning}
\end{figure}

In Algorithm \ref{alg:active_learning}, at each step after initialization, the model can query one extensometer or one convergence sensor. We study two possibilities: the model can query i) only extensometer measurements or ii) only convergence measurements. At each step, when new sensors are queried, the corresponding collocation, boundary, and data points are simultaneously added to the training set. This ensures that the data, PDE, and boundary condition loss terms are all evaluated for the newly included samples during subsequent training. The proposed algorithms are greedy in nature, meaning that they do not guarantee convergence toward a global optimum. Nevertheless, increasing the number of accurate observations generally provides more information to the model, which helps constrain the solution space and solve the inverse problem.  Figure \ref{fig:active_learning} illustrates the active learning procedure presented in Algorithm \ref{alg:active_learning}.

Note that in Algorithm~\ref{alg:active_learning}, the model is first initialized using two extensometers before the full grid is added. This strategy ensured that the model converged towards the true solution. One extensometer is placed at the crown and another on the left side of the tunnel, providing the initial set of observational data. This choice is motivated by the fact that these locations typically exhibit the most significant deformations in tunneling, making them natural starting points for model calibration. Additional observational data are then progressively incorporated when the corresponding measurement points are selected by AL.

\section{Results}\label{sec:results}

To evaluate the accuracy of the predicted displacement field $\boldsymbol{\hat{u}}$ in reference to the true displacement field $\boldsymbol{u}$, we compute the relative $L^2$ error over the test set:
\begin{equation*}
\frac{\lVert \boldsymbol{\hat{u}} - \boldsymbol{u} \rVert_{2}}{\lVert \boldsymbol{u} \rVert_{2}}.
\end{equation*}

For the inverse problem, the accuracy of the estimated rock mass parameter $\hat{\lambda}$ is quantified using the relative parameter error:
\begin{equation*}
\frac{\lvert \hat{\lambda} - \lambda \rvert}{\lvert \lambda \rvert}.
\end{equation*}
where $\lambda$ denotes a true parameter. 

To account for randomness arising from sampling, network initialization, and the optimization process, each experiment is repeated 10 times using random seeds, and the mean and standard deviation of the errors are reported. 

\subsection{Case study}

To illustrate the approach, we aim to simultaneously reconstruct the displacement field and back-calculate rock properties around the Saint-Martin-la-Porte access gallery, located in Savoie, France. For context, the Saint-Martin-la-Porte access gallery was excavated between 2003 and 2010. During excavation, the rock mass displayed pronounced anisotropic behavior, primarily associated with the Carboniferous formation (also known as the Productive Houillier). 

In the following, we generate the synthetic dataset using the parameters obtained by Vu et al. \cite{vu_semi-analytical_2013}: the elastic parameters are $E_h = 620$ MPa, $E_v = 340$ MPa, $G_{vh} = 200$ MPa, $\nu_h = 0.12$, and $\nu_{hv} = 0.2$; the vertical stress is assumed to be $5$ MPa, with a lateral pressure coefficient $K = 0.75$; and a varying bedding plane angle $\beta \in \{0^\circ, 45^\circ, 90^\circ, 135^\circ$\}. Here, we aim to infer $E_h$, $E_v$, $G_{vh}$, $\beta$, and $K$ while $\nu_h$, and $\nu_{hv}$ remain fixed. It is assumed that the tunnel is fully excavated, such that the radial stress at the wall is zero.

\subsection{Reconstruction of the displacement field}

\begin{figure}[H]
\centering
\begin{subfigure}[b]{0.48\textwidth}
    \centering
    \includegraphics[scale=1]{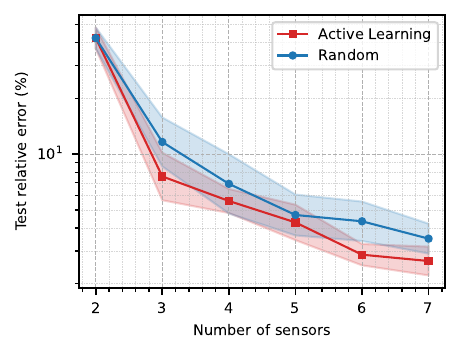}
    \caption{Querying only extensometers.}
    \label{fig:random_vs_al_ext}
\end{subfigure}
\begin{subfigure}[b]{0.48\textwidth}
    \centering
    \includegraphics[scale=1]{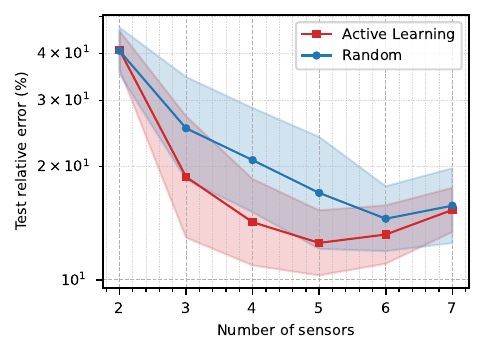}
    \caption{Querying only convergence measurements.}
    \label{fig:random_vs_al_conv}
\end{subfigure}
\caption{Comparison of relative test errors obtained using active learning versus random selection for different types of sensors. The solid lines correspond to the mean values of the errors, and the shaded regions represent the associated standard deviations. Obtained for ten different random seeds in the case of $\beta = 45^\circ$.}
\label{fig:random_vs_al}
\end{figure}

We first compare the active learning strategy based on uncertainty-driven sampling against a random selection method. The evolution of the relative error, evaluated on the test set, is shown in Figure \ref{fig:random_vs_al}. Two querying methods are compared: sampling from extensometer data only (Figure~\ref{fig:random_vs_al_ext}) and sampling from convergence data only (Figure~\ref{fig:random_vs_al_conv}). 

As shown in Figure~\ref{fig:random_vs_al}, selecting measurements based on the highest estimated uncertainty consistently improves model performance compared to random sampling for both querying strategies. The performance gain is most pronounced when only a few measurements are available, with an improvement of 6--7\% using three extensometers and 2--3\% using four extensometers (Figure~\ref{fig:random_vs_al_ext}). This effect is more significant when only convergence measurements are queried, as illustrated in Figure~\ref{fig:random_vs_al_conv}. For a given query strategy (extensometer only or convergence only), the difference between random and active learning selection decreases as more measurements are included, since additional data help the model reconstruct the displacement field in regions with the largest uncertainties. This confirms the usefulness of AL in cases where observation data are scarce.

\begin{figure}[H]
\centering
\begin{subfigure}{0.47\linewidth}
\centering
\includegraphics[scale=1]{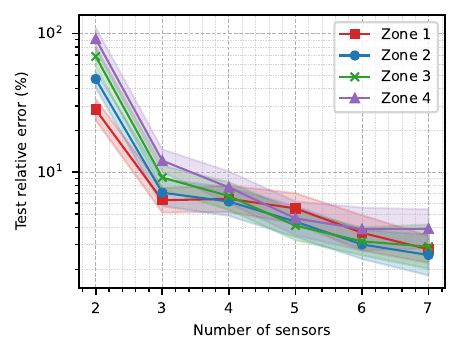}
\caption{$u_x$ querying extensometers}
\label{fig:ux_ext}
\end{subfigure}
\begin{subfigure}{0.47\linewidth}
\centering
\includegraphics[scale=1]{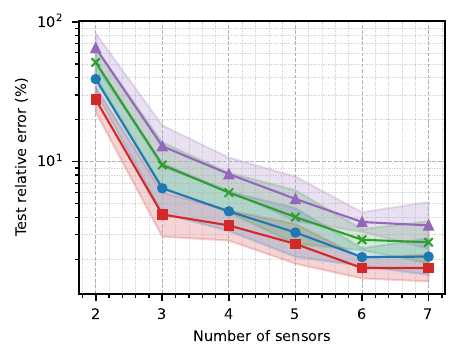}
\caption{$u_y$ querying extensometers}
\label{fig:uy_ext}
\end{subfigure}
\centering
\begin{subfigure}{0.47\linewidth}
\centering
\includegraphics[scale=1]{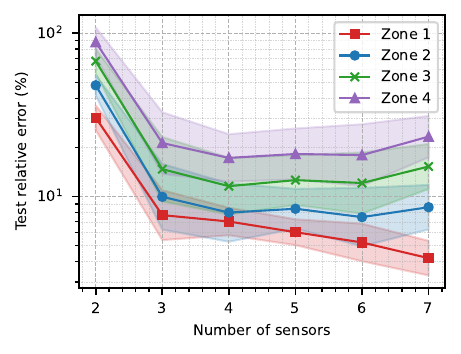}
\caption{$u_x$ querying convergences}
\label{fig:ux_cv}
\end{subfigure}
\begin{subfigure}{0.47\linewidth}
\centering
\includegraphics[scale=1]{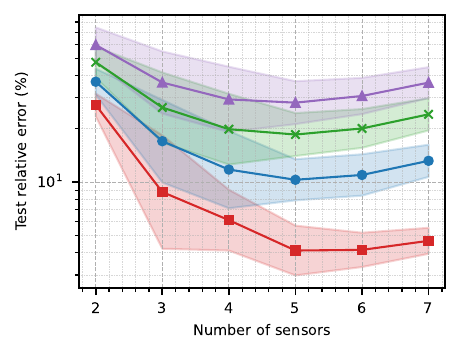}
\caption{$u_y$ querying convergences}
\label{fig:uy_cv}
\end{subfigure}
\caption{Mean relative test error evaluated in distinct zones. Zone 1: $[R, 2R]$, Zone 2: $[2R, 3R]$, Zone 3: $[3R, 4R]$, Zone 4: $[4R, 5R]$. Obtained for ten different random seeds in the case of $\beta = 45^\circ$.}
\label{fig:al1_vs_al2}
\end{figure}

Figure \ref{fig:al1_vs_al2} compares the model performance on the test set, which is divided into four concentric zones around the tunnel wall. Zone 1 corresponds to points within the interval $[R, 2R]$, Zone 2 to $[2R, 3R]$, Zone 3 to $[3R, 4R]$, and Zone 4 to $[4R, 5R]$. Figures \ref{fig:ux_ext} and \ref{fig:uy_ext} show the model performance for the horizontal and vertical displacement components, $u_x$ and $u_y$, respectively, when only extensometer data are queried. Similarly, Figures \ref{fig:ux_cv} and \ref{fig:uy_cv} present the results for the horizontal and vertical displacement components when only convergence data are queried. For both querying strategies, the active learning selection method enables the model to progressively improve its performance as more measurements are queried. Overall, querying extensometer data yields lower relative test errors compared to querying only convergence data.

In Zone 1, both querying strategies achieve similar performance. This suggests that it is possible in practice to reconstruct the displacement field around the tunnel with high accuracy up to one tunnel radius, with only two extensometers, those used for initialization at the crown and at the waist, and convergence data. With three extensometers, the error is approximately 4–5\% for $u_y$, while with the two initial extensometer data and one queried convergence, the error on $u_y$ is 7–8\%. These errors gradually decrease to 3\% and 5\%, respectively, when five sensors in total are used. In Zones 2, 3, and 4, querying extensometer data yields better accuracy than querying convergence data: errors generally remain below 10\% when using three extensometers and below 6\% with five extensometers. By contrast, querying only convergence data does not sufficiently reduce the prediction error in Zones 2, 3, and 4, which are located farther from the tunnel wall. This outcome was expected, as no additional measurements from these zones are included in the training set.

\begin{figure}[h]
    \centering
    \begin{subfigure}{0.4\linewidth}
        \centering
        \includegraphics[scale=1]{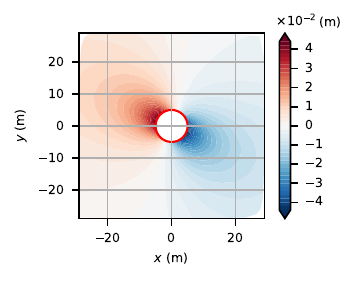}
        \centering
        \caption{$u_x$ true}
    \end{subfigure}
    \centering
    \begin{subfigure}{0.4\linewidth}
        \centering
        \includegraphics[scale=1]{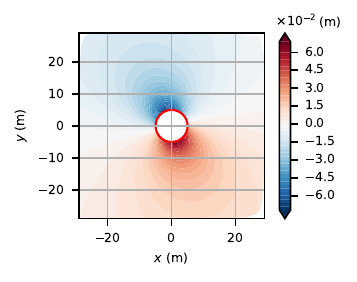}
        \centering
        \caption{$u_y$ true}
    \end{subfigure}
    \caption{Computations made using the MOOSE framework.}
    \label{fig:true_data_moose}
\end{figure}

\begin{figure}[h]
    \begin{subfigure}{0.32\linewidth}
        \centering
        \includegraphics[scale=0.94]{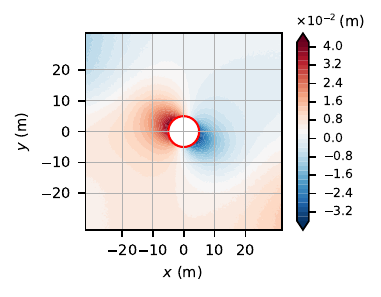}
        \centering
        \caption{2 extensometers: $u_x$}
    \end{subfigure}
    \centering
    \begin{subfigure}{0.32\linewidth}
        \centering
        \includegraphics[scale=0.94]{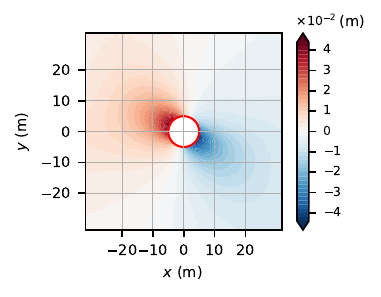}
        \centering
        \caption{3 extensometers: $u_x$}
    \end{subfigure}
    \begin{subfigure}{0.32\linewidth}
        \centering
        \includegraphics[scale=0.94]{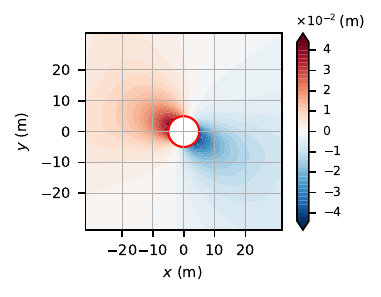}
        \centering
        \caption{4 extensometers: $u_x$}
    \end{subfigure}
    \begin{subfigure}{0.32\linewidth}
        \centering
        \includegraphics[scale=0.94]{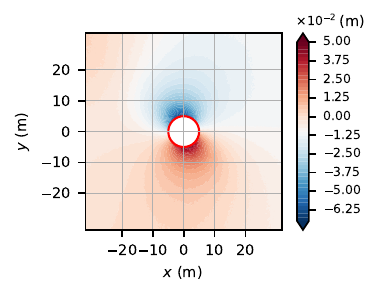}
        \centering
        \caption{2 extensometers: $u_y$}
    \end{subfigure}
    \centering
    \begin{subfigure}{0.32\linewidth}
        \centering
        \includegraphics[scale=0.94]{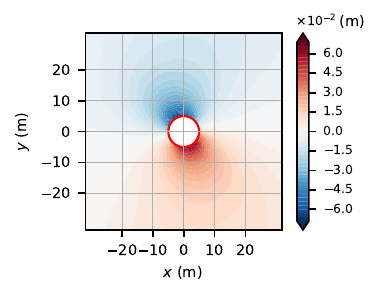}
        \centering
        \caption{3 extensometers: $u_y$}
    \end{subfigure}
    \begin{subfigure}{0.32\linewidth}
        \centering
        \includegraphics[scale=0.94]{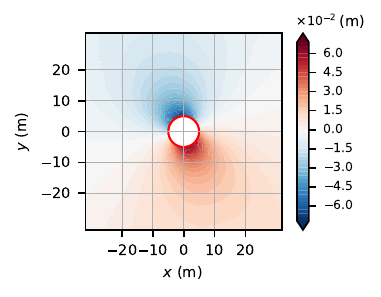}
        \centering
        \caption{4 extensometers: $u_y$}
    \end{subfigure}
    \caption{Sequential active training and progressive reconstruction of the displacement field when querying only extensometer data. Obtained in the case of $\beta = 45^\circ$.}
    \label{fig:al_1_disp}
\end{figure}

\begin{figure}[H]
    \begin{subfigure}{0.32\linewidth}
        \centering
        \includegraphics[scale=0.94]{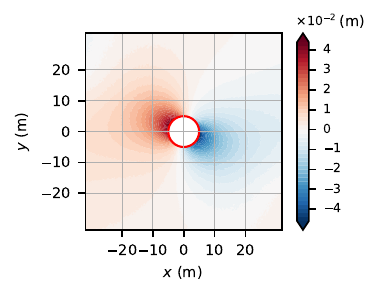}
        \centering
        \caption{1 convergence queried: $u_x$}
    \end{subfigure}
    \centering
    \begin{subfigure}{0.32\linewidth}
        \centering
        \includegraphics[scale=0.94]{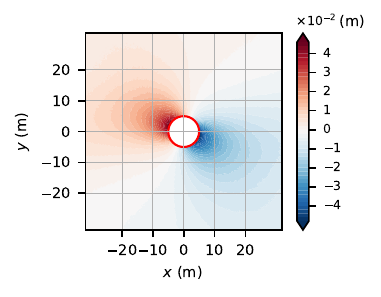}
        \centering
        \caption{2 convergences queried: $u_x$}
    \end{subfigure}
    \begin{subfigure}{0.32\linewidth}
        \centering
        \includegraphics[scale=0.94]{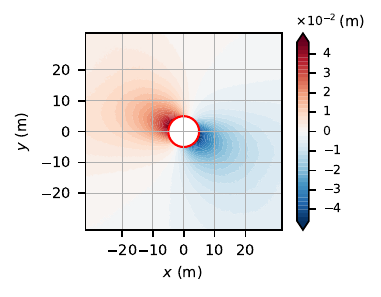}
        \centering
        \caption{3 convergences queried: $u_x$}
    \end{subfigure}
    \begin{subfigure}{0.32\linewidth}
        \centering
        \includegraphics[scale=0.94]{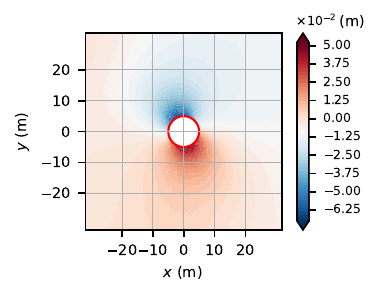}
        \centering
        \caption{1 convergence queried: $u_y$}
    \end{subfigure}
    \centering
    \begin{subfigure}{0.32\linewidth}
        \centering
        \includegraphics[scale=0.94]{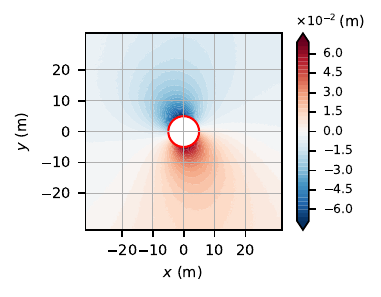}
        \centering
        \caption{2 convergences queried: $u_y$}
    \end{subfigure}
    \begin{subfigure}{0.32\linewidth}
        \centering
        \includegraphics[scale=0.94]{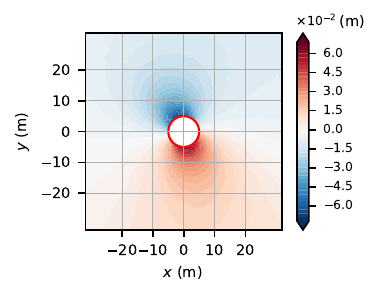}
        \centering
        \caption{3 convergences queried: $u_y$}
    \end{subfigure}
    \caption{Sequential active training and progressive reconstruction of the displacement field when querying only convergence data. Obtained in the case of $\beta = 45^\circ$.}
    \label{fig:al_2_disp}
\end{figure}

Figure~\ref{fig:true_data_moose} shows the true displacement field generated using the finite element software \textsc{MOOSE}. Figures~\ref{fig:al_1_disp} and \ref{fig:al_2_disp} illustrate the model predictions when only extensometer data are queried and when only convergence data are queried, respectively. The PINNs are trained using the sequential active learning procedures described in Section~\ref{sec:AL}. The corresponding displacement predictions along the tunnel wall are shown for several training steps, after which no further changes are observed. To remain consistent with the available data, the predictions are restricted to a maximum radial distance of $d = R + L$, beyond which no measurements are assumed to be available.

In Figure~\ref{fig:al_1_disp}, the displacement field is partially reconstructed by the model using only two extensometers. With three or more extensometers, the displacement field is fully reconstructed for both horizontal and vertical components. Overall, the accuracy of the predicted displacement fields, both near the cavity and farther away, improves as additional extensometers are included. Similarly, in Figure~\ref{fig:al_2_disp}, a progressive reconstruction of the displacement field is observed when only convergence data are queried. Using two extensometers and one convergence sensor, the model already reconstructs both displacement components around the tunnel with good accuracy. When two or more convergence measurements are queried, the model accurately reconstructs the displacement field close to the tunnel wall, although the predictions are less accurate in Zones 2, 3, and 4, which are farther from the wall.

Table~\ref{tab:displacement_errors_combined} reports the relative errors on $u_x$ and $u_y$ for $\beta \in \{0^\circ°, 45^\circ, 90^\circ, 135^\circ\}$ and both querying strategies, using 5 sensors in total.

\begin{table}[H]
\centering
\caption{Relative error (\%) on the displacement components $u_x$ and $u_y$ around the tunnel for different bedding plane orientations $\beta$ and radial zones. Extensometers: 3 queried (5 in total). Convergences: 3 queried (2 extensometers + 3 convergences in total). Zone 1: $[R, 2R]$, Zone 2: $[2R, 3R]$, Zone 3: $[3R, 4R]$, Zone 4: $[4R, 5R]$.}
\label{tab:displacement_errors_combined}
\resizebox{\textwidth}{!}{%
\renewcommand{\arraystretch}{1.2}
\begin{tabular}{llcccccccc}
\toprule
& & \multicolumn{2}{c}{$\beta = 0°$} 
  & \multicolumn{2}{c}{$\beta = 45°$} 
  & \multicolumn{2}{c}{$\beta = 90°$} 
  & \multicolumn{2}{c}{$\beta = 135°$} \\
\cmidrule(lr){3-4}\cmidrule(lr){5-6}\cmidrule(lr){7-8}\cmidrule(lr){9-10}
Comp. & Zone
  & Extenso. & Converg. 
  & Extenso. & Converg. 
  & Extenso. & Converg. 
  & Extenso. & Converg. \\
\midrule
\multirow{4}{*}{$u_x$}
& 1 & $6.2 \pm 0.9$   & $9.3 \pm 1.4$ 
    & $5.7 \pm 1.4$   & $6.1 \pm 1.1$  
    & $2.1 \pm 0.3$   & $4.9 \pm 0.8$  
    & $4.7 \pm 0.7$   & $7.4 \pm 1.4$  \\
& 2 & $10.6 \pm 2.7$  & $17.2 \pm 1.9$ 
    & $4.6 \pm 1.2$   & $8.7 \pm 2.2$  
    & $3.1 \pm 0.9$   & $12.5 \pm 2.4$ 
    & $3.8 \pm 0.9$   & $10.2 \pm 2.6$ \\
& 3 & $13.6 \pm 3.2$  & $21.5 \pm 3.4$ 
    & $4.2 \pm 1.0$   & $13.4 \pm 4.4$ 
    & $4.4 \pm 1.5$   & $21.5 \pm 4.9$ 
    & $4.6 \pm 1.4$   & $15.0 \pm 6.1$ \\
& 4 & $16.8 \pm 3.9$  & $24.9 \pm 5.9$ 
    & $4.8 \pm 1.2$   & $19.4 \pm 6.7$  
    & $6.2 \pm 2.1$   & $31.4 \pm 7.8$  
    & $6.7 \pm 2.4$   & $20.9 \pm 9.5$  \\
\midrule
\multirow{4}{*}{$u_y$}
& 1 & $2.2 \pm 0.7$   & $6.8 \pm 1.7$  
    & $2.7 \pm 1.0$   & $4.3 \pm 1.2$  
    & $3.4 \pm 0.8$   & $6.8 \pm 1.9$  
    & $3.1 \pm 0.9$   & $6.7 \pm 2.5$  \\
& 2 & $3.0 \pm 1.2$   & $18.5 \pm 3.3$ 
    & $3.4 \pm 1.7$   & $10.7 \pm 2.8$ 
    & $5.5 \pm 1.7$   & $12.7 \pm 3.8$ 
    & $3.4 \pm 1.0$   & $15.0 \pm 5.2$ \\
& 3 & $4.1 \pm 1.3$   & $32.7 \pm 5.1$ 
    & $4.5 \pm 2.4$   & $19.3 \pm 5.4$ 
    & $7.0 \pm 2.2$   & $18.8 \pm 6.7$ 
    & $4.5 \pm 1.1$   & $24.6 \pm 7.8$ \\
& 4 & $5.0 \pm 1.3$   & $47.7 \pm 6.8$ 
    & $5.8 \pm 2.7$   & $29.2 \pm 8.1$ 
    & $8.4 \pm 2.7$   & $26.4 \pm 9.8$ 
    & $6.0 \pm 1.4$   & $35.1 \pm 10.2$ \\
\bottomrule
\end{tabular}}
\end{table}

From Table \ref{tab:displacement_errors_combined}, it can be observed that, when only extensometer data are queried (``extensometer mode''), low relative errors are achieved across all zones for $\beta \in \{45^\circ, 90^\circ, 135^\circ\}$. In particular, errors remain below approximately 5--7\% in Zone~1 for both $u_x$ and $u_y$, and increase moderately with distance, staying below 10\% overall. For $\beta = 0^\circ$, the model exhibits larger errors in the prediction of $u_x$. This can be explained by the fact that the horizontal displacement component has a low amplitude in this configuration, making its reconstruction more difficult. In contrast, $u_y$, which dominates the displacement field in this case, is predicted with good accuracy. When only convergence data are queried after the two initial extensometer data are read (``convergence mode''), the relative errors are systematically higher than in extensometer mode for all zones and bedding plane orientations. Nevertheless, the displacement field remains reasonably well reconstructed in Zone 1, while the accuracy degrades more significantly when the distance from the tunnel increases.

Finally, these results indicate that the proposed active learning strategy effectively enables (i) the selection of the most informative measurements to enhance model performance, and (ii) the progressive reconstruction of the displacement field around a deep circular tunnel. 

A limitation of the proposed approach is that a minimum of two extensometers is required to initialize the active learning process. If fewer measurements are employed, the model fails to reconstruct the displacement field and to identify the physical parameters. Furthermore, the current algorithm is inherently greedy, as it selects only the next measurement point without considering the overall arrangement of sensors. As a result, it does not optimize a global sensing strategy and may miss combinations of points that could collectively provide more informative or efficient coverage of the domain.

\subsection{Inverse analysis}

We now evaluate the results of the inverse analysis. The corresponding results, reported in Table \ref{tab:inverse_beta}, show the estimated parameters at step 5 of the active learning process for both extensometer and convergence modes obtained after conducting for each case ten simulations using different random seeds.

As shown in Table \ref{tab:inverse_beta}, the model accurately recovers the physical parameters in both extensometer and convergence modes, although the latter exhibits higher errors overall. A notable observation is that the relative errors on $E_h$ (respectively $E_v$) are larger for $\beta = 0^\circ$ (respectively $\beta = 90^\circ$). This behavior can be explained by the reduced sensitivity of the displacement field to these parameters in the corresponding configurations. When the bedding planes are horizontal (respectively vertical), the displacement field is dominated by vertical (respectively horizontal) displacements, primarily governed by $E_v$ (respectively $E_h$). As a result, $E_h$ (respectively $E_v$) has a weaker influence on the observable response, leading to reduced identifiability in the inverse problem.

\begin{table}[h!]
\centering
\caption{Relative error (\%) on the optimized model parameters obtained at step 5 of the active learning process. Results obtained after conducting ten computations corresponding to ten different random seeds.}
\label{tab:inverse_beta}
\resizebox{\textwidth}{!}{%
\renewcommand{\arraystretch}{1.2}
\begin{tabular}{lcccccccc}
\toprule
& \multicolumn{2}{c}{$\beta = 0°$} 
  & \multicolumn{2}{c}{$\beta = 45°$} 
  & \multicolumn{2}{c}{$\beta = 90°$} 
  & \multicolumn{2}{c}{$\beta = 135°$} \\
\cmidrule(lr){2-3}\cmidrule(lr){4-5}\cmidrule(lr){6-7}\cmidrule(lr){8-9}
Parameter 
  & Extenso. & Converg. 
  & Extenso. & Converg. 
  & Extenso. & Converg. 
  & Extenso. & Converg. \\
\midrule
$E_h$    & $19.4 \pm 7.8$  & $23.0 \pm 7.1$  
         & $2.7 \pm 2.2$  & $3.6 \pm 2.7$  
         & $5.7 \pm 2.2$  & $5.9 \pm 4.5$  
         & $4.5 \pm 2.5$  & $6.3 \pm 3.8$  \\
$E_v$    & $4.0 \pm 1.5 $ & $3.8 \pm 2.5$  
         & $5.0 \pm 3.7$  & $5.6 \pm 2.9$  
         & $16.5 \pm 6.6$  & $33.0 \pm 8.5$  
         & $1.4 \pm 1.1$  & $8.0 \pm 5.6$  \\
$G_{vh}$ & $6.0 \pm 2.0$  & $6.7 \pm 2.6$  
         & $2.4 \pm 1.3$  & $4.3 \pm 3.4$  
         & $1.2 \pm 1.0$  & $4.4 \pm 2.5$  
         & $4.2 \pm 2.1$  & $7.4 \pm 4.3$  \\
$K$      & $8.6 \pm 2.7$  & $8.9 \pm 2.8$  
         & $1.2 \pm 0.8$  & $1.1 \pm 0.9$  
         & $8.3 \pm 2.9$  & $11.9 \pm 3.9$  
         & $1.0 \pm 0.8$  & $5.7 \pm 3.1$  \\
$\beta$  & $1.3 \pm 1.0$  & $4.8 \pm 2.0$  
         & $5.0 \pm 4.6$  & $9.5 \pm 5.3$  
         & $1.1 \pm 0.8$  & $6.1 \pm 3.6$  
         & $1.1 \pm 0.9$  & $7.4 \pm 4.4$  \\
\bottomrule
\end{tabular}}
\end{table}

\begin{figure}[H]
    \centering
    \begin{subfigure}{0.47\linewidth}
        \centering
        \includegraphics[scale=1]{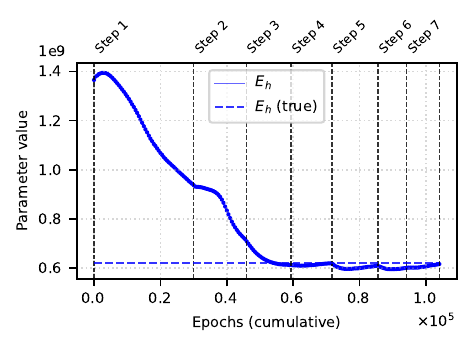}
        \centering
        \caption{}
    \end{subfigure}
    \centering
    \begin{subfigure}{0.47\linewidth}
        \centering
        \includegraphics[scale=1]{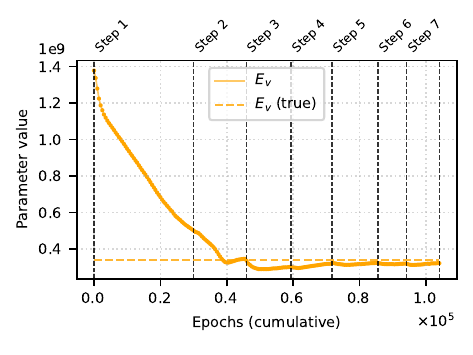}
        \centering
        \caption{}
    \end{subfigure}
    \begin{subfigure}{0.47\linewidth}
        \centering
        \includegraphics[scale=1]{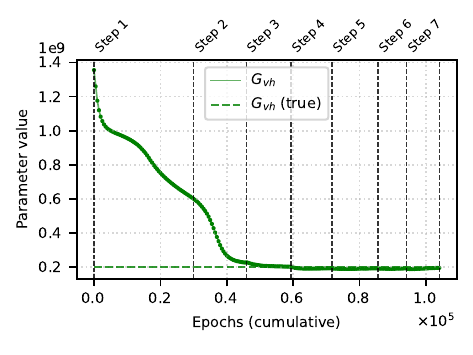}
        \centering
        \caption{}
    \end{subfigure}
    \centering
    \begin{subfigure}{0.47\linewidth}
        \centering
        \includegraphics[scale=1]{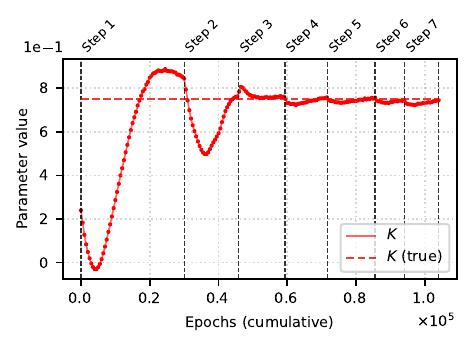}
        \centering
        \caption{}
    \end{subfigure}
    \centering
    \begin{subfigure}{0.47\linewidth}
        \centering
        \includegraphics[scale=1]{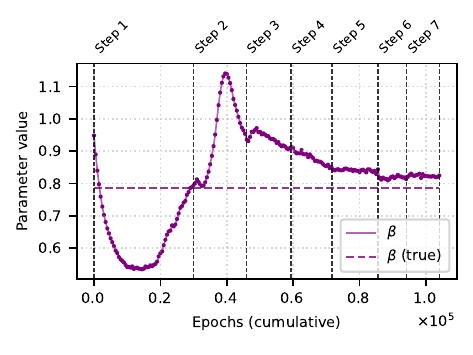}
        \centering
        \caption{}
    \end{subfigure}
    \caption{Sequential optimization of parameters using active training. Vertical dashed lines denote the stages at which additional measurements are queried and incorporated into the training set. Obtained in the case of $\beta = 45^\circ$.} 
    \label{fig:params_optim}
\end{figure}

To illustrate the optimization process during training, Figure \ref{fig:params_optim} shows the values of the parameters optimized by the PINN as a function of the number of training epochs when using the active learning strategy with extensometer data only. Step 1 and step 2 correspond to model initialization, as detailed in Algorithm \ref{alg:active_learning}. In step 1, the training set consists of the observations from the two initial extensometers and their associated collocation/boundary points. In Step 2, the grid $\mathcal{G}$ is added to the training set. At each step, optimization begins with a reinitialized optimizer having a learning rate of $l_r = 10^{-4}$, and the full batch of training data, that is, all available training points in $\mathcal{T}$, is provided to the model. Early stopping is applied, resulting in a different number of epochs at each stage.

\begin{figure}[h]
    \centering
    \begin{subfigure}{0.47\linewidth}
        \centering
        \includegraphics[scale=1]{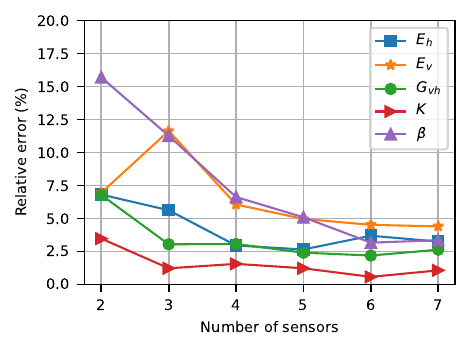}
        \centering
        \caption{Querying only extensometers.}
    \end{subfigure}
    \centering
    \begin{subfigure}{0.47\linewidth}
        \centering
        \includegraphics[scale=1]{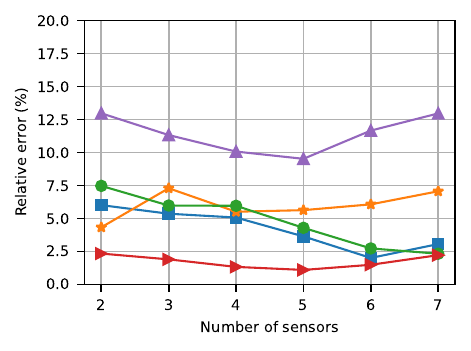}
        \centering
        \caption{Querying only convergences.}
    \end{subfigure}
    \caption{Constitutive parameters obtained by the model after optimization step by step. The curves correspond to mean results obtained over 10 different seeds in the case of $\beta = 45^\circ$.} 
    \label{fig:mean_relative_errors_params}
\end{figure}

To illustrate the sequential approach, Figure \ref{fig:mean_relative_errors_params} presents the mean relative errors obtained with Algorithm \ref{alg:active_learning} when querying either extensometer data or convergence data exclusively in the case of $\beta = 45^\circ$. It can be observed that the sought parameters, i.e., \(E_h\), \(E_v\), \(G_{vh}\), \(K\), and \(\beta\), are successfully determined under both querying strategies. When querying extensometer data, $E_h$, $G_{vh}$, and $K$ are the most accurately estimated parameters, with errors below 8\% using two extensometers and below 5\% with three. In contrast, the errors for $E_v$ and $\beta$ remain under 15\% with two and three extensometers, and decrease to below 6\% and 5\%, respectively, once four or more extensometers are used.  

A similar trend is observed when querying only convergence data. The errors for $E_h$, $G_{vh}$, and $K$ fall below 5\% after three convergence measurements are added to the data pool, and continue to decrease thereafter. The error for $E_v$ remains below 10\% and gradually reduces to approximately 6\% with five convergence measurements. The error associated with $\beta$ remains between 10\% and 15\% across all steps. This result can be attributed to the fact that convergence data are concentrated along the tunnel wall, which limits the model’s ability to accurately optimize $\beta$, a parameter primarily influencing the displacement orientation rather than local displacement magnitudes.  

\subsection{Computational time} 

Table \ref{tab:timing} reports the computational time per active learning step, averaged over 10 seeds on an Apple M4 silicon chip. For both extensometer and convergence modes, each step requires between 3 and 6 minutes, and the full active learning process, from initialization to 7 sensors, is completed in approximately 25 minutes. This demonstrates that the proposed framework is computationally efficient and compatible with real-time tunnel monitoring, where excavation steps typically last from several hours to several days, leaving sufficient time to query new sensors, retrain the model, and update the displacement field and constitutive parameter estimates before the next excavation round.

\begin{table}[h!]
\centering
\renewcommand{\arraystretch}{1.2}
\caption{Computational time per active learning step and cumulative on an Apple M4 silicon chip. Step 1 and step 2 correspond to the initialization with 2 extensometers, common to both querying strategies.}
\label{tab:timing}
\begin{tabular}{cccccc}
\toprule
& & \multicolumn{2}{c}{Extensometers} & \multicolumn{2}{c}{Convergences} \\
\cmidrule(lr){3-4}\cmidrule(lr){5-6}
Step & Sensors & Time (s) & Cumul. (s) & Time (s) & Cumul. (s) \\
\midrule
1 & 2 & $185 \pm 5$ & $185 \pm 5$ & $189 \pm 5$ & $189 \pm 5$ \\
2 & 2 & $305 \pm 26$ & $490 \pm 27$ & $286 \pm 36$ & $475 \pm 37$ \\
3 & 3 & $246 \pm 32$ & $736 \pm 47$ & $246 \pm 35$ & $721 \pm 58$ \\
4 & 4 & $224 \pm 24$ & $960 \pm 49$ & $195 \pm 31$ & $916 \pm 77$ \\
5 & 5 & $209 \pm 32$ & $1169 \pm 62$ & $178 \pm 34$ & $1094 \pm 80$ \\
6 & 6 & $190 \pm 24$ & $1359 \pm 53$ & $158 \pm 34$ & $1252 \pm 103$ \\
7 & 7 & $191 \pm 31$ & $1550 \pm 60$ & $167 \pm 26$ & $1419 \pm 110$ \\
\bottomrule
\end{tabular}
\end{table}

\subsection{Model behavior using noisy data}

We now analyze the sensitivity of the model to noisy data as it is an important problem in tunnelling where in-situ measurements are often noisy \cite{guayacan2026lessons}. To assess the robustness of the model, synthetic heteroscedastic Gaussian noise was added to the displacement data. The noisy data $\tilde{\mathbf{u}}$ were generated as:
\begin{equation}
\tilde{\mathbf{u}} = \mathbf{u} + \eta, \qquad 
\eta \sim \mathcal{N}\left(0, \alpha^2 \mathbf{u}^2 \right),
\end{equation}
where $\mathbf{u}$ denotes the true displacement vector and $\alpha$ controls the noise amplitude. In this work, $\alpha$ was set to 0.05, 0.10, and 0.15 to simulate noise levels of 5\%, 10\%, and 15\%, respectively.

\begin{figure}[H]
\centering
\begin{subfigure}{0.48\textwidth}
    \centering
    \includegraphics[width=\textwidth]{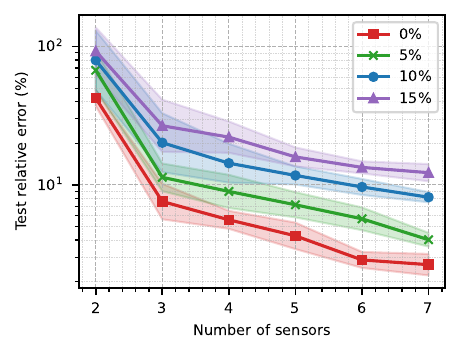}
    \centering
    \caption{Query on extensometers only.}
    \label{fig:noise_extensometers}
\end{subfigure}
\centering
\begin{subfigure}{0.48\textwidth}
    \centering
    \includegraphics[width=\textwidth]{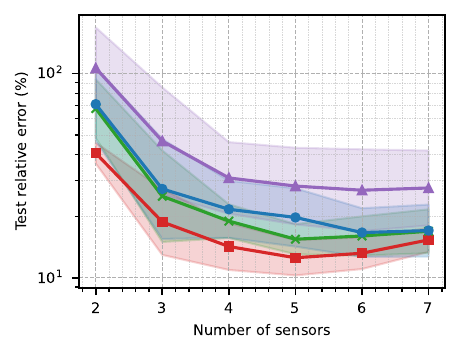}
    \centering
    \caption{Query on convergence only.}
    \label{fig:noise_cv}
\end{subfigure}
\caption{Comparison of relative error in the test set for different noise magnitudes. The solid lines correspond to the mean values of the errors obtained over 10 repeated runs with different seeds, and the shaded regions represent the associated standard deviations.}
\label{fig:noise_rel_err}
\end{figure}

Figure~\ref{fig:noise_rel_err} shows the relative displacement errors evaluated on the test set for different noise magnitudes. For noise levels of 5\% and 10\%, the model still achieves a good accuracy for both querying strategies. Using three extensometers, the error is approximately equal to 10\% (respectively, 20\%) for  a noise level of 5\% (respectively, 10\%). With five extensometers, this error drops to 7\% (respectively, 10\%) for  a noise level of 5\% (respectively, 10\%). At 15\% of noise, the performance decreases, with errors around 25\% at five extensometers. Furthermore, the standard deviation of the predictions increases, indicating a reduced confidence. Using two extensometer data and one convergence measurement, the error in the whole domain is about 25 \% for 5\% and 10\% of noise magnitudes. This error decreases to approximately 17-20\% when three convergence measurements are queried. At 15\% of noise, the error is around 30\% when three convergences are queried. For both querying strategies, a noise level above 15\% considerably decreases model performance, at which point, the model can no longer converge. Further investigations are needed to treat inverse problems with high levels of noise - for example, through a Bayesian perspective \cite{stuart_inverse_2010}.

\begin{figure}[H]
    \centering
    \begin{subfigure}{0.47\linewidth}
        \centering
        \includegraphics[width=\linewidth]{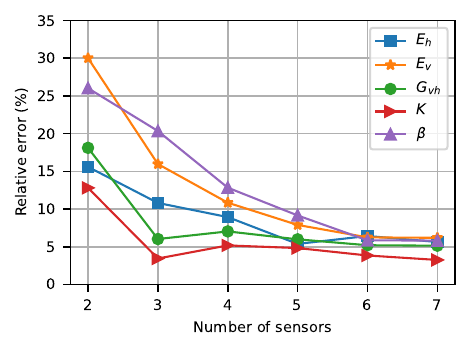}
        \caption{Noise 5\% querying extensometers}
        \label{fig:noise_5}
    \end{subfigure}
    \begin{subfigure}{0.47\linewidth}
        \centering
        \includegraphics[width=\linewidth]{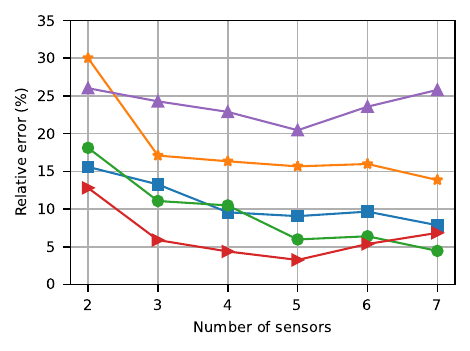}
        \caption{Noise 5\% querying convergences}
        \label{fig:noise_5_cv}
    \end{subfigure}
    \begin{subfigure}{0.47\linewidth}
        \centering
        \includegraphics[width=\linewidth]{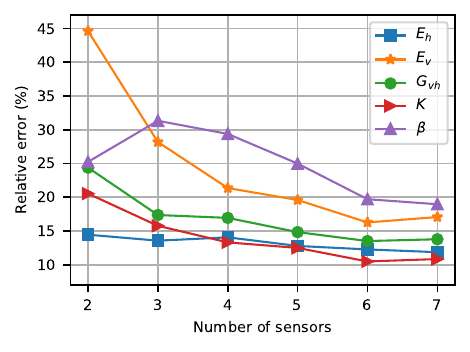}
        \caption{Noise 10\% querying extensometers}
        \label{fig:noise_10}
    \end{subfigure}
    \begin{subfigure}{0.47\linewidth}
        \centering
        \includegraphics[width=\linewidth]{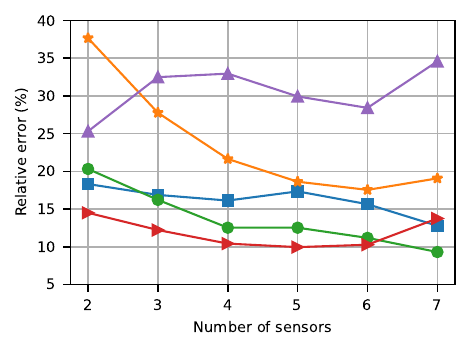}
        \caption{Noise 10\% querying convergences}
        \label{fig:noise_10_cv}
    \end{subfigure}
    \begin{subfigure}{0.47\linewidth}
        \centering
        \includegraphics[width=\linewidth]{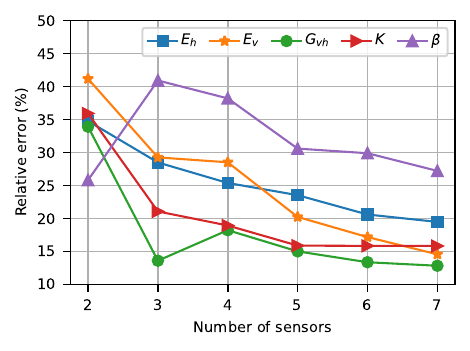}
        \caption{Noise 15\% querying extensometers}
        \label{fig:noise_15}
    \end{subfigure}
    \begin{subfigure}{0.47\linewidth}
        \centering
        \includegraphics[width=\linewidth]{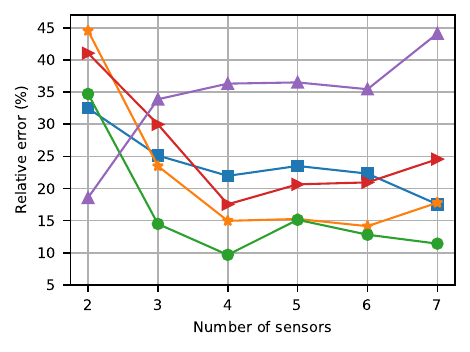}
        \caption{Noise 15\% querying convergences}
        \label{fig:noise_15_cv}
    \end{subfigure}
    \caption{Results of the inverse analysis for multiple noise levels, comparing extensometer and convergence data querying strategies. Solid lines indicate the mean errors obtained over 10 repeated runs with different seeds. Standard deviations are not represented for clarity.}
    \label{fig:mean_relative_errors_params_noise}
\end{figure}

Figure~\ref{fig:mean_relative_errors_params_noise} presents the results of the inverse analysis for noisy measurements with amplitudes of 5\%, 10\%, and 15\%. The model provides reasonable predictions with data that contains small to moderate levels of noise: the overall trends are consistent with those obtained from noise-free data, although the mean relative error increases with noise level. For 5\% noise, the model achieves errors below 15\% for \(E_h\), \(G_{vh}\), and \(K\) using three extensometers, which progressively improves to approximately 5\% with five extensometers (Figure~\ref{fig:noise_5}). The parameters \(E_v\) and \(\beta\) are also correctly estimated, with errors around 17\% and 25\% at three extensometers, and errors of 10\% at five extensometers. At a noise level of 10\%, a similar trend is observed. Errors for \(E_h\), \(G_{vh}\), and \(K\) range between 15\% and 18\% for three extensometers and decrease to 13\%-15\% with five extensometers (Figure~\ref{fig:noise_10}). For \(E_v\) and \(\beta\), the errors are approximately 30\% at three extensometers and reduce to around 20\% and 25\%, respectively, at five extensometers. For 15\% noise, optimization becomes more challenging. The relative errors for \(E_h\), \(E_v\), \(G_{vh}\), and \(K\) lie between 15\% and 25\% with five extensometers, while the error for \(\beta\) remains relatively high, close to 30\% (Figure~\ref{fig:noise_15}). When querying only convergence data, similar observations can be made although the direction of the bedding planes $\beta$ is more difficult to determine.

\section{Conclusion}\label{sec:Conclusion}

In this study, we propose a methodology to reconstruct the displacement field and characterize the rock mass surrounding deep circular tunnels excavated in transversely isotropic elastic media. To this end, we leverage the mesh-free nature of PINNs to back-analyze extensometer and convergence measurements collected during tunneling. 

The main contribution of this work lies in enabling the PINN to autonomously select informative displacement measurements through an active learning framework based on epistemic uncertainty, quantified through dropout regularization. This strategy allows the sequential selection of sensors among extensometers and convergence measurements, which reduces the number of required observations and optimizes sensor placement. The developed methodology ultimately improves the reliability of the inverse identification and accuracy of the reconstructed displacement field.

More specifically, by incorporating equilibrium equations, anisotropic constitutive laws, and boundary conditions as soft constraints into the loss function, the model can reconstruct the displacement field and identify the elastic parameters of the rock mass, the horizontal in-situ stress state, and the orientation of the planes of isotropy, even when observations are limited, irregularly distributed, or affected by small to moderate noise levels.

Lastly, the methodology enables back-analysis using data collected at the tunnel scale during the excavation phase. The proposed tool is compatible with traditional observational methods as it can support near real time decision-making for optimal monitoring of ground behavior. Future work will focus on improved sensor-deployment strategies based on active planning computational methods. Other research directions include time-dependent effects typical of squeezing conditions and non-circular sections.

\section{Code and data availability}
The code and data used in this manuscript are publicly available at the GitHub repository \cite{tristani2026_AL_Tunnel_code} (see also \url{https://github.com/Alec-YT/AL-Tunnel}). 

\section{Acknowledgments}
The authors wish to thank Prof. Herbert Einstein from MIT for discussions on sensor positioning and Prof. Yunan Yang from Cornell for her advice on training optimization.

\appendix

\section{Sensitivity analysis of loss weights}
\label{app:sensitivity}

 This section presents a sensitivity analysis of the loss weights $w_{\mathrm{bc}}$ and $w_{\mathrm{data}}$ with $w_{\mathrm{pde}} = 1$ fixed, conducted over the full active learning process (up to step 5) in extensometer mode and convergence  mode for $\beta = 135^\circ$. 
 
 Following the observations of Wang et al. \cite{wang_understanding_2020, wang_when_2022}, that imbalanced gradients across loss terms can lead to training instabilities, we explored different weights for the data and boundary loss terms. In particular, the weights were increased as powers of 10 and 9 configurations were tested in total. Each of the 9 configurations was repeated ten times using different random seeds, corresponding to different network parameter initializations, resulting in 90 training runs for each extensometer and convergence modes (180 computations in total). 
 
As reported in Table \ref{tab:sensitivity_convergence}, convergence is reached if a relative test error lower than $30\%$ is achieved; otherwise, the model is declared non-convergent.
\begin{table}[h!]
\centering
\renewcommand{\arraystretch}{1.2}
\caption{Convergence of the active learning procedure for different loss weight configurations ($w_{\mathrm{pde}} = 1$ fixed, extensometer mode (Extenso.) and convergence mode (Converg.), $\beta = 135°$). "y" indicates convergence over 10 seeds; "n" indicates no convergence.}
\label{tab:sensitivity_convergence}
\begin{tabular}{cccc}
\toprule
$w_{\mathrm{bc}}$ & $w_{\mathrm{data}}$ & Extenso. & Converg. \\
\midrule
$1$   & $1$    & n & n \\
$1$   & $10$   & n & n \\
$1$   & $100$  & n & n \\
$1$   & $1000$ & n & n \\
$10$  & $10$   & y & y \\
$10$  & $100$  & y & y \\
$10$  & $1000$ & n & n \\
$100$ & $100$  & y & y \\
$100$ & $1000$ & y & y \\
\bottomrule
\end{tabular}
\end{table}

From table \ref{tab:sensitivity_convergence}, it appears that the model is able to converge toward the right solution when $w_{\mathrm{data}} \geq w_{\mathrm{bc}} > w_{\mathrm{pde}}$. When $w_{\mathrm{bc}} = w_{\mathrm{pde}}= 1$, the model fails to reconstruct the displacement field as the boundary conditions, which explicitly contain the parameters $K$ and $\beta$, are insufficiently enforced. Conversely, when $w_{\mathrm{data}}$ is excessively large, the data gradient dominates, and the PDE residuals are neglected, leading to a physically inconsistent solution.

\begin{figure}[h]
\centering
\begin{subfigure}{0.48\textwidth}
    \centering
    \includegraphics[width=\textwidth]{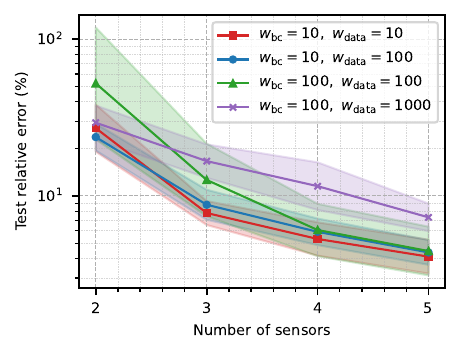}
    \centering
    \caption{Query on extensometers only.}
    \label{fig:noise_extensometers}
\end{subfigure}
\centering
\begin{subfigure}{0.49\textwidth}
    \centering
    \includegraphics[width=\textwidth]{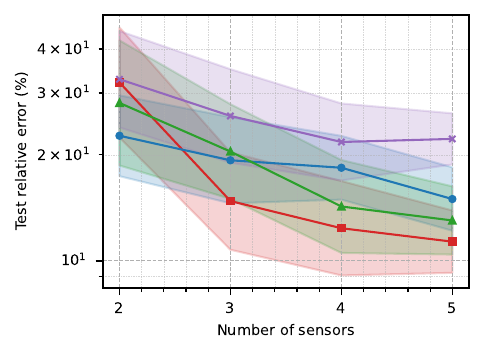}
    \centering
    \caption{Query on convergence only.}
    \label{fig:noise_cv}
\end{subfigure}
\caption{Sensitivity analysis of loss weights in extensometer mode 
($\beta = 135^\circ$, $w_{\mathrm{pde}} = 1$ fixed). 
Results are averaged over 10 seeds and shaded regions represent one standard deviation.}
\label{fig:sensitivity}
\end{figure}

Figure \ref{fig:sensitivity} shows the evolution of the relative test error as a function of the number of sensors queried during the active learning process for the four configurations that yield convergence.

\bibliographystyle{unsrtnat} 
\bibliography{bibliography.bib}

\end{document}